\documentclass[floatfix,reprint,aps,twocolumn,amssymb,superscriptaddress,showkeys]{revtex4-2}

\usepackage{mathtools} 
\usepackage{graphicx}
\usepackage{epstopdf}
\usepackage{amsmath}
\usepackage{dcolumn}
\usepackage{bm}
\usepackage{bbold}
\usepackage{amssymb}
\usepackage{mathrsfs}
\usepackage{amsfonts}
\usepackage{braket}
\usepackage{tabularx}
\usepackage{cleveref}
\usepackage{setspace}
\usepackage[caption=false]{subfig}

\usepackage{float}
\usepackage[toc,page]{appendix}
\usepackage{array}
\usepackage[dvipsnames]{xcolor}
\usepackage{enumitem}
\usepackage{comment}
\usepackage{multirow}
\usepackage{makecell}
\usepackage{comment}
\usepackage{tikz}

\newcolumntype{L}[1]{>{\raggedright\let\newline\\\arraybackslash\hspace{0pt}}m{#1}}
\newcolumntype{C}[1]{>{\centering\let\newline\\\arraybackslash\hspace{0pt}}m{#1}}
\newcolumntype{R}[1]{>{\raggedleft\let\newline\\\arraybackslash\hspace{0pt}}m{#1}}

\newcommand{\figref}[1]{Fig.~\ref{#1}}
\newcommand{\secref}[1]{Sec.~\ref{#1}}
\newcommand{\appref}[1]{Appendix~\ref{#1}}
\newcommand{\tabref}[1]{table~\ref{#1}}
\renewcommand{\eqref}[1]{eq.~(\ref{#1})}

\newcommand{\citeall}[1]{\citeauthor{#1} (\citeyear{#1}) \cite{#1}}

\begin{document}

\title{Level attraction in a quasi-closed cavity}

\author{Guillaume Bourcin}
\email{guillaume.bourcin@imt-atlantique.fr}
\affiliation{IMT Atlantique, Technopole Brest-Iroise, CS 83818, 29238 Brest Cedex 3, France}
\affiliation{Lab-STICC (UMR 6285), CNRS, Technopole Brest-Iroise, CS 83818, 29238 Brest Cedex 3, France}
\author{Alan Gardin}
\affiliation{IMT Atlantique, Technopole Brest-Iroise, CS 83818, 29238 Brest Cedex 3, France}
\affiliation{Lab-STICC (UMR 6285), CNRS, Technopole Brest-Iroise, CS 83818, 29238 Brest Cedex 3, France}
\affiliation{School of Physics, The University of Adelaide, Adelaide SA 5005, Australia}
\author{Jeremy Bourhill}
\affiliation{Quantum Technologies and Dark Matter Labs, Department of Physics,  University of Western Australia, 35 Stirling Hwy, 6009 Crawley, Western Australia.\\
}%
\author{Vincent Vlaminck}
\affiliation{IMT Atlantique, Technopole Brest-Iroise, CS 83818, 29238 Brest Cedex 3, France}
\affiliation{Lab-STICC (UMR 6285), CNRS, Technopole Brest-Iroise, CS 83818, 29238 Brest Cedex 3, France}
\author{Vincent Castel}
\email{vincent.castel@imt-atlantique.fr}
\affiliation{IMT Atlantique, Technopole Brest-Iroise, CS 83818, 29238 Brest Cedex 3, France}
\affiliation{Lab-STICC (UMR 6285), CNRS, Technopole Brest-Iroise, CS 83818, 29238 Brest Cedex 3, France}

\date{\today}

\begin{abstract}
    We provide a comprehensive analytical description of the effective coupling associated with an antiresonance within a hybrid system comprised of a quasi-closed photonic cavity and a ferrimagnetic material. Whilst so-called level attraction between a resonant system inside an open cavity is well understood, the physical underpinnings of this phenomena within quasi-closed cavities have remained elusive. Leveraging the input-output theory, we successfully differentiate between the repulsive and attractive aspects of this coupling. Our proposed model demonstrates that by understanding the phase-jump at the resonances and the studied antiresonance, we can predict the nature of the effective coupling of the antiresonance for a given position of the ferrimagnet in the cavity.
\end{abstract}

\maketitle

\printfigures
\section{Introduction}

Cavity spintronics is an expanding field of research dedicated to the precise control and exploration of the interactions between light and magnetic materials. Notably, the interaction between cavity photons and magnons, the quanta of spin waves, has gained significant prominence since 2010 \cite{barman2021MagnonicsRoadmap2021,lachance-quirionHybridQuantumSystems2019,zarerameshtiCavityMagnonics2022}.
When a magnon is tuned coincident in frequency with a photonic resonance, and the coupling rate between the two systems is faster than their individual loss mechanisms, the two separate systems become hybridised into what is referred to as a cavity magnon polariton (CMP).

In cavity magnonics, two distinct types of resonators exist \cite{zarerameshtiCavityMagnonics2022}: open cavities, characterized by an antiresonance (dip) in the transmission spectra at their mode frequencies, and quasi-closed cavities, identified by a resonance (peak) in the transmission at their mode frequencies. While only the level repulsion is possible between a cavity resonance and a magnon mode, the effective interaction between a cavity antiresonance and a magnon mode can lead to either level repulsion or level attraction, which is characterized by a crossing or anticrossing in the dispersion spectrum, respectively \cite{hardercoherentdissipativeCavity2021,wangdissipativeCouplingsCavity2020,zarerameshtiCavityMagnonics2022}. The observation of level attraction in cavity magnonics systems was first reported by \citeall{harderLevelAttractionDue2018}. 
The physical processes governing level attraction in open cavities has been explained by employing traveling waves \cite{bhoiAbnormalAnticrossingEffect2019,harderLevelAttractionDue2018,raoInteractionsMagnonMode2020,yaocoherentControlMagnon2019,yaoMicroscopicOriginMagnonphoton2019,yangControlMagnonphotonLevel2019}.

In contrast, the attractive character of the coupling in a quasi-closed cavity was solely experimentally identified by \citeall{raoLevelAttractionLevel2019}. To explain this observation, a phenomenological model based on RLC circuits was employed. However, this approach did not provide a clear understanding of the mechanisms underlying the emergence of level attraction.
A more comprehensive understanding of the origins of this phenomenon is a valuable insight for the design of cavities in various applications, such as magnon-magnon interactions \cite{grigoryanCavitymediateddissipativeSpinspin2019,reiterScalabledissipativePreparation2016,xuCavitymediateddissipativeCoupling2019}, or non-reciprocal wave propagation \cite{barzanjehMechanicalOnchipMicrowave2017,bernierNonreciprocalReconfigurableMicrowave2017,calozElectromagneticNonreciprocity2018,lecocqNonreciprocalMicrowaveSignal2017,metelmannNonreciprocalPhotonTransmission2015,petersonDemonstrationEfficientNonreciprocity2017} (e.g. backscattering isolation or unidirectional signal amplification).

In this paper, we focus on the coupling between an antiresonance and a magnon mode by applying input-output theory. This approach provides a deeper understanding of the parameters that govern the coupling behavior, whether it is repulsive or attractive. We begin by introducing the general $S$-matrix derived from input-output theory and proceed to explore the two distinct pathways leading to the occurrence of antiresonance.
This analysis provides significant insight into the factors contributing to the antiresonance coupling phenomenon, particularly focusing on the phase-jump, which is the key feature of the effective antiresonance coupling behaviour.

Finally, the agreement between finite element method (FEM) simulations and our proposed model prove the possibility to precisely control the coupling behavior when positioning a ferrimagnetic sphere at different locations in a quasi-closed cavity.


\section{Physical Model}

\begin{figure*}[t!]
    \centering
    \includegraphics[width=\textwidth]{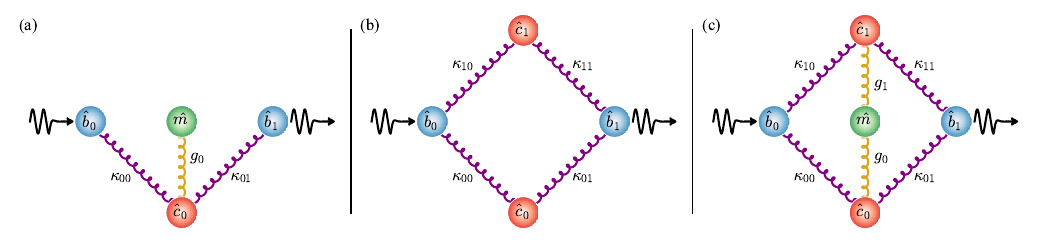}
    \captionsetup{width=\textwidth}
    \vspace{-1cm}
    \caption{Schemes of classical oscillators representing a quasi-closed cavity composed of (a) a photonic mode $\hat{c}_0$ coupled with strength $g_0$ to a magnon mode $\hat{m}$; (b) two photonic modes $\hat{c}_0$ and $\hat{c}_1$; and (c) two photonic modes $\hat{c}_0$ and $\hat{c}_1$ coupled with strength $g_0$ and $g_1$ respectively to a magnon mode $\hat{m}$. $\hat{b}_0$ and $\hat{b}_1$ are the ports, and $\kappa_{ij} = \sqrt{\gamma_{ij}} e^{i \phi_{ij}}$ the external couplings.}
    \label{fig:systems}
\end{figure*}

The model Hamiltonian for a closed cavity, denoted as $\hat{\mathcal{H}}_\mathrm{sys}$, encompasses $p$ internal bosonic modes, which can be either photons or magnons in the context of this study. The Hamiltonian is expressed as follows:
\begin{equation}
    \dfrac{\hat{\mathcal{H}}_\mathrm{sys}}{\hbar} = \sum_{p}{
        \left[\tilde{\omega}_p \hat{a}_p^\dagger(t) \hat{a}_p(t) + \dfrac{1}{2} \sum_{q \neq p}{
            \left(g_{qp} \hat{a}_p^\dagger(t) \hat{a}_q(t) + \mathrm{h.c.}\right)
        }\right]
    }.
\end{equation}
The first term represents the unperturbed Hamiltonian of a single oscillator, where $\omega_p/2\pi$ is the eigenfrequency, and $\hat{a}_p^\dagger$ ($\hat{a}_p$) is the creation (annihilation) operator of mode $p$. The second term is the interaction Hamiltonian between two internal modes $\hat{a}_p$ and $\hat{a}_q$, with their mutual coupling assessed by $g_{qp}$, and h.c. indicating the hermitian conjugate. It is worth noting that the fast oscillating terms (i.e. $\hat{a}_p^\dagger\hat{a}_q^\dagger$ and $\hat{a}_p\hat{a}_q$) are neglected in this approximation, which is known as the Rotating Wave Approximation (RWA) \cite{leboiteTheoreticalMethodsUltrastrong2020}.

In a quasi-closed cavity, the cavity modes are coupled to a common photon bath \cite{gardinerInputOutputDamped1985,yuanLoopTheoryInputoutput2020}. For each port $n$ (i.e. a probe), an associated photon bath is represented by a continuum of photonic oscillators, named external modes, with eigenfrequency $\omega$. The related Hamiltonian can be expressed as follow:
\begin{equation}
    \label{eq:Ham_bath}
    \dfrac{\hat{\mathcal{H}}_\mathrm{bath}}{\hbar} = \sum_n{
    \int_\mathbb{R}{
        \mathrm{d}\omega \, \omega \hat{b}_{\omega, n}^\dagger(t) \hat{b}_{\omega, n}(t)
    }
    }.
\end{equation}
Here, $\hat{b}_{\omega, n}^\dagger$ ($\hat{b}_{\omega, n}$) represents the creation (annihilation) operator of the external mode associated with port $n$ and having the frequency $\omega$.

The interaction between the bath and the system can be described by the following model Hamiltonian (under the RWA):
\begin{equation}
    \label{eq:Ham_int}
    \dfrac{\hat{\mathcal{H}}_\mathrm{int}}{\hbar} = \dfrac{i}{\sqrt{2\pi}} \sum_{p, n}{
        \int_\mathbb{R}{
            \mathrm{d}\omega \left(
                \kappa_{pn}(\omega) \hat{b}_{\omega, n}^\dagger(t) \hat{a}_p(t) - \mathrm{h.c.}
            \right)
        }
    },
\end{equation}
where $\kappa_{pn}(\omega)$ is the external coupling strength between the external mode $\hat{b}_{n, \omega}$ and the internal mode $\hat{a}_p$.

In the first Markov approximation, the external coupling strength is assumed to be independent of the frequency:
\begin{equation}
    \label{eq:Markov}
    \kappa_{pn}(\omega) = \kappa_{pn} = \sqrt{\gamma_{pn}} e^{i \phi_{pn}}, \quad \gamma_{pn} \in \mathbb{R}.
\end{equation}
Here, $\gamma_{pn}$ is real and represents the external photonic damping rate. Additionally, a phase contribution $\phi_{pn}$ is introduced to the external coupling $\kappa_{pn}$, as previously discussed in \cite{bourhillGenerationCirculatingCavity2023,zhangBroadbandNonreciprocityEnabled2020}.
The coupling phase is contingent on the phase of the electric (magnetic) field injected or probed within the cavity. At a given time $t$, the first probe injects a field with a certain phase. Consequently, each excited cavity mode shares the same phase at this location in their field distribution within the cavity. However, the phase of the modes at the second probe may differ, according to their field distribution, giving rise to both constructive and destructive interferences. A probe senses the field only along one axis, thereby resulting in an external coupling phase of either $0$ or $\pi$.

The complete Hamiltonian is given by:
\begin{equation}
    \label{eq:Hamiltonian}
    \hat{\mathcal{H}} = \hat{\mathcal{H}}_{\mathrm{sys}} + \hat{\mathcal{H}}_\mathrm{bath} + \hat{\mathcal{H}}_{\mathrm{int}}.
\end{equation}
It is important to note that, owing to the RWA, this Hamiltonian is no longer applicable to systems operating beyond the Strong-Coupling (SC) regime, where $g_{qp}/\sqrt{\omega_q \omega_p} < 0.1$ and $\kappa_{pn}(\omega)/\sqrt{\omega_p} < 0.1$ \cite{ciutiInputoutputTheoryCavities2006,forn-diazUltrastrongCouplingRegimes2019,friskkockumUltrastrongCouplingLight2019}.
Both internal and external modes obey to the bosonic relation \footnote[3]{%
\begin{minipage}{\linewidth}
  \begin{flalign*}
    &\begin{aligned}
      &[\hat{a}_i(t), \hat{a}_j^\dagger(t)] = \delta_{ij}, \\
      &[\hat{b}_{\omega, i}(t), \hat{b}_{\omega', j}^\dagger(t)] = \delta_{ij}\delta(\omega - \omega'),
    \end{aligned}&
  \end{flalign*}
\end{minipage}
where $\delta_{ij}$ and $\delta(\omega - \omega')$ are the Kronecker symbols.
}.
The derivation of the $S$-matrix using the input-output theory is given in \appref{sec:theory}.

\section{Physics of an antiresonance}
\label{sec:antires}

The following section will discuss the two contributions to the existence of an antiresonance with two internal modes: when both modes are hybridized Cavity-Magnon Polaritons (CMP), and when both modes are photonic modes. These scenarios are illustrated in the two system schemes in \figref{fig:systems} (a), and (b), respectively. \figref{fig:systems} (c), a combination of \figref{fig:systems} (a) and (b), will be discussed in \secref{sec:coupling}.

\subsection{One photon mode \& one magnon mode}
\label{subsec:photon_magnon}
\begin{figure}[b!]
    \centering
    \includegraphics[width=\linewidth]{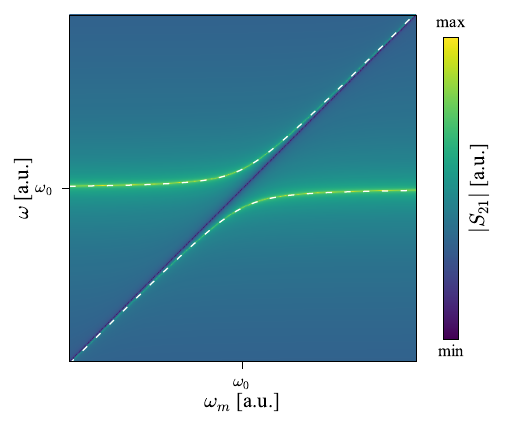}
    \vspace{-1cm}
    \caption{Transmission spectrum of a system composed of a cavity photon and a magnon, sketched in \figref{fig:systems} (a), using \eqref{eq:photon_magnon}, and the polariton frequency dependence in dashed white line using \eqref{eq:polaritons}.}
    \label{fig:magnon_photon}
\end{figure}

The depicted system in \figref{fig:systems} (a) consists of a single photonic mode $\hat{c}_0$ with a frequency of $\omega_0$, interacting with two ports (i.e. two probes) denoted as $\hat{b}_0$ and $\hat{b}_1$. The coupling strengths for these interactions are characterized by $\kappa_{00}$ and $\kappa_{01}$, respectively. Additionally, the magnon $\hat{m}$ with a frequency of $\omega_m$ is only coupled to the photonic mode with a coupling strength of $g_0$.
According to \eqref{eq:QLEsol} and \eqref{eq:S}, the transmission of this system is expressed as follows:
\begin{equation}
    \label{eq:photon_magnon}
    S_{21} = -i \dfrac{\sqrt{\gamma_{00} \gamma_{01}} \Delta_m e^{i \Phi_0}}{\tilde{\Delta}_0 \Delta_m - g_0^2},
\end{equation}
where $\Delta_m = \omega - \omega_m$, $\tilde{\Delta}_0 = \omega - \tilde{\omega}_0$, $\tilde{\omega}_0 = \omega_0 - \dfrac{i}{2} (\gamma_{00} + \gamma_{01})$, and $\Phi_0 = \phi_{01} - \phi_{00}$.\\
Minimizing the denominator in \eqref{eq:photon_magnon} gives rise to the polaritonic frequencies:
\begin{equation}
    \label{eq:polaritons}
    \omega_\pm = \dfrac{1}{2} \left[\tilde{\omega}_0 + \omega_m \pm \sqrt{(\omega_m - \tilde{\omega}_0)^2 + 4g_0^2}\right].
\end{equation}
Minimizing the numerator of \eqref{eq:photon_magnon} leads to the determination of the antiresonance frequency $\omega_\mathrm{ar}$ arising from the interaction between a photon and a magnon:
\begin{equation}
    \omega_{ar} = \omega_m.
\end{equation}
For this case, regardless of the coupling phase between the cavity photon and the ports, the antiresonance frequency will always coincide with the magnon frequency, which is the externally non-excited oscillator, as highlighted in the transmission spectra of the system depicted in \figref{fig:magnon_photon}.

Note that \eqref{eq:photon_magnon} can be expressed as a sum of Lorentzians:
\begin{equation}
    \label{eq:photon_magnon_lor}
    S_{21} = -i \dfrac{\sqrt{\gamma_{00}\gamma_{01}}}{\omega_+ - \omega_-} \left(\dfrac{\omega_m - \omega_-}{\omega - \omega_-} + \dfrac{\omega_+ - \omega_m}{\omega - \omega_+}\right) e^{i\Phi_0}.
\end{equation}
The terms within the parentheses correspond to the resonances of the lower and upper polaritons. Numerators and the shared factor term (except for $e^{i \Phi_0}$) are all positive. As a result, the phase of both resonances shift from $\pi/2$ to $-\pi/2$ when $\Phi_0 = 0$ (referred to as negative phase-jump in all subsequent discussions), and from $-\pi/2$ to $\pi/2$ when $\Phi_0 = \pi$ (positive phase-jump). The common phase factor $e^{i \Phi_0}$ for both polaritons indicates that these resonances undergo the same phase-jump.

To emphasize the phase-jump of the antiresonance, the transmission can be rewritten in a different manner:
\begin{equation}
    \label{eq:Spolaritons}
    S_{21} = -i \dfrac{\sqrt{\gamma_{00} \gamma_{01}} \Delta_m e^{i\Phi_0}}{(\omega - \omega_-)(\omega - \omega_+)}.
\end{equation}
In a frequency range between the two polariton frequencies, the given equation characterizes the antiresonance. In this frequency range, all terms maintain the same sign except for $\Delta_m$, which influences the phase-jump. To ensure a meaningful comparison of the phase in \eqref{eq:photon_magnon_lor}, it is crucial that all terms remain positive (excluding $\Delta_m$ and $e^{i \Phi_0}$); otherwise, the phase-jump would be affected. However, $\omega - \omega_+$ is negative around the antiresonance frequency. Therefore, to render all terms positive in the expression of the transmission, \eqref{eq:Spolaritons} becomes:
\begin{equation}
    S_{21} = -i \dfrac{\sqrt{\gamma_{00} \gamma_{01}} \Delta_m e^{i\Phi_\mathrm{ar}}}{(\omega - \omega_-)(\omega_+ - \omega)},
\end{equation}
where $\Phi_\mathrm{ar} = \Phi_0 + \pi$ represents the phase factor of the antiresonance, and its phase factor is $\pi$-dephased compared to the phase factors of the polaritons, meaning that the phase-jump is positive when $\Phi_0 = 0$, and negative when $\Phi = \pi$.
Further discussion of this phase signature will be presented in the following sections.

\subsection{Two photon modes}
\label{subsec:2photons}
The system illustrated in \figref{fig:systems} (b) comprises two photonic modes $\hat{c}_0$ and $\hat{c}_1$ with frequencies $\omega_0$ and $\omega_1$, respectively. These two photon modes interact with two ports, $\hat{b}_0$ and $\hat{b}_1$, characterized by external coupling strengths denoted as $\kappa_{00}$ and $\kappa_{01}$ with respect to mode $\hat{c}_0$, and $\kappa_{10}$ and $\kappa_{11}$ with respect to mode $\hat{c}_1$.
From \eqref{eq:QLEsol} and \eqref{eq:S}, the transmission of such a system reads:
\begin{equation}
    \label{eq:2photons}
    S_{21} = -i \dfrac{\sqrt{\gamma_{00} \gamma_{01}} \tilde{\Delta}_1  e^{i \Phi_0} + \sqrt{\gamma_{10} \gamma_{11}} \tilde{\Delta}_0 e^{i \Phi_1} + \dfrac{i}{2} \Gamma_1}{\tilde{\Delta}_0 \tilde{\Delta}_1 + \dfrac{|\Gamma|}{4}},
\end{equation}
where $\tilde{\Delta}_1 = \omega - \tilde{\omega}_1$, $\tilde{\omega}_1 = \omega_1 - \dfrac{i}{2} (\gamma_{10} + \gamma_{11})$, $\Phi_1 = \phi_{11} - \phi_{10}$, $\Gamma = \sqrt{\gamma_{00} \gamma_{10}} e^{i(\phi_{10} - \phi_{00})} + \sqrt{\gamma_{01} \gamma_{11}} e^{i(\phi_{11} - \phi_{01})}$, and $\Gamma_1 = \sqrt{\gamma_{00} \gamma_{11}} \Gamma e^{i(\phi_{11} - \phi_{00})} - \sqrt{\gamma_{01} \gamma_{10}} \Gamma^* e^{i(\phi_{10} - \phi_{01})}$.\\
The resonance frequencies of the transmission reads:
\begin{equation}
    \omega_\pm = \dfrac{1}{2} \left[\tilde{\omega}_0 + \tilde{\omega}_1 \pm \sqrt{(\tilde{\omega}_1 - \tilde{\omega}_0)^2 - |\Gamma|}\right].
\end{equation}
In the context of a cavity exhibiting non-degenerate modes, the following hypothesis applies:
\begin{equation}
    \label{eq:hypothesis}
    |\omega_1 - \omega_0| \gg |\Gamma|.
\end{equation}
This inequality reflects the condition that the frequency difference between the two modes is significantly larger than the magnitude of $\Gamma$, interpreted as a crosstalk between the two photonic modes. In this approximation, the resonance frequencies are equal to the photon mode frequencies:
\begin{equation}
    \omega_- = \tilde{\omega}_0, \qquad \omega_+ = \tilde{\omega}_1.
\end{equation}
The antiresonance frequency of such a system is given by:
\begin{equation}
    \label{eq:antires}
    \omega_\mathrm{ar} = \dfrac{\omega_1 + \delta e^{i \Phi} \omega_0}{1 + \delta e^{i \Phi}},
\end{equation}
where $\delta = \sqrt{\gamma_{10}\gamma_{11}/\gamma_{00}\gamma_{01}}$ represents the external dissipation ratio between the mode $\hat{c}_1$ and $\hat{c}_0$ at the two probes, and $\Phi = \Phi_1 - \Phi_0 = \phi_{00} + \phi_{11} - \phi_{01} - \phi_{10}$ represents the phase factor difference between the two photonic modes.

We highlight three distinct antiresonance frequency regimes based on the values of $\delta$ and $\Phi$, summarized in \tabref{tab:antiresonance}, and discussed just below.
\begin{table}[b!]
    \caption{Antiresonance behavior}
    \begin{ruledtabular}
        \label{tab:antiresonance}%
        \begin{tabular}{c|cc}
            & $\bm{\Phi = 0}$ & $\bm{\Phi = \pi}$ \\
            \hline
            $\bm{\delta > 1}$ & \multirow{2}{*}{$\omega_\mathrm{ar} \in [\omega_0; \omega_1]$} & $\omega_\mathrm{ar} \in [0; \omega_0]$\\
            $\bm{\delta < 1}$ &  & $\omega_\mathrm{ar} \in [\omega_1; +\infty[$\\
        \end{tabular}
    \end{ruledtabular}
\end{table}

Considering the hypothesis given in \eqref{eq:hypothesis}, the transmission from \eqref{eq:2photons} can also be expressed as a sum of Lorentzians:
\begin{equation}
    \label{eq:s21photon}
    S_{21} = -i \left(\dfrac{\sqrt{\gamma_{00}^* \gamma_{01}}  e^{i \Phi_0} - i \Gamma_2}{\omega - \tilde{\omega}_0} + \dfrac{\sqrt{\gamma_{10}^* \gamma_{11}}  e^{i \Phi_1} - i \Gamma_2}{\omega - \tilde{\omega}_1}\right),
\end{equation}
where $\Gamma_2 = \dfrac{\Gamma^*}{\tilde{\omega}_1 - \tilde{\omega}_0} \Gamma_1$, is negligible far from the antiresonance frequency. By neglecting $\Gamma_2$, the two terms represent isolated photon modes.
\begin{figure}[b!]
    \centering
    \includegraphics[width=\linewidth]{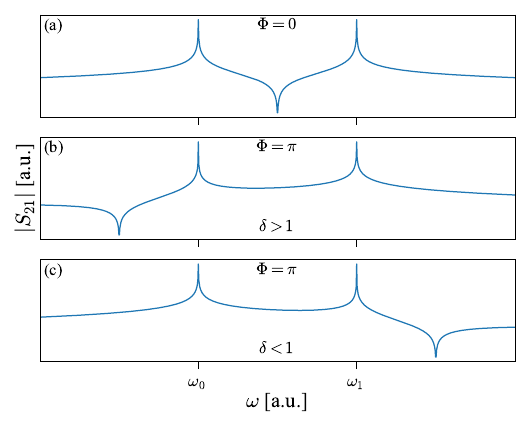}
    \vspace{-1cm}
    \caption{Three cases of the frequency dependence of the antiresonance of the system composed of two cavity photons, sketched in \figref{fig:systems} (b). $\delta$ and $\Phi$ are defined in \eqref{eq:antires}.}
    \label{fig:2photons}
\end{figure}

When $\Phi=0$, the antiresonance frequency $\omega_\mathrm{ar}$ is situated between the two photon mode frequencies $\omega_0$ and $\omega_1$, as depicted in \figref{fig:2photons} (a). This corresponds to the scenario where the two Lorentzians in \eqref{eq:s21photon} share the same phase factor, i.e. $\Phi_0 = \Phi_1$, therefore the same phase-jump. Thereby, the two Lorentzian are $\pi$-dephased and destruct themselves for frequencies between $\omega_0$ and $\omega_1$. This aligns with the case discussed in \secref{subsec:photon_magnon} in \eqref{eq:photon_magnon_lor}; when two eigenmodes share the phase factor, the antiresonance frequency lies between their respective eigenfrequencies.

When $\Phi=\pi$, the antiresonance frequency $\omega_\mathrm{ar}$ lies outside the two photon mode frequencies $\omega_0$ and $\omega_1$. This corresponds to the scenario where the phase factor of the two Lorentzians in \eqref{eq:s21photon} are $\pi$-dephased, therefore opposed phase-jumps. Two cases arise: when $\delta > 1$, indicating that the photon mode $\hat{c}_1$ is more coupled with the probes than the photon mode $\hat{c}_0$, and the antiresonance frequency $\omega_\mathrm{ar}$ is lower than $\omega_0$, as illustrated in \figref{fig:2photons} (b); when $\delta < 1$, indicating that the photon mode $\hat{c}_0$ is more coupled with the probes than the photon mode $\hat{c}_1$, and $\omega_\mathrm{ar}$ is higher than $\omega_1$, as shown in \figref{fig:2photons} (c).

As in the previous section, the transmission can be rewritten to emphasize the phase of the antiresonance:
\begin{equation}
    S_{21} = -i \dfrac{|1 + \delta|(\omega - \omega_\mathrm{ar})e^{i \Phi_\mathrm{ar}} + \dfrac{i}{2}\Gamma_3}{(\omega - \omega_-)(\omega - \omega_+)},
\end{equation}
where $\Gamma_3 = \sqrt{\gamma_{00}\gamma_{01}}(\gamma_{10} + \gamma_{11}) + \sqrt{\gamma_{10}\gamma_{11}}(\gamma_{00} + \gamma_{01}) + \Gamma_1$, is negligible far from the antiresonance frequency, and $\Phi_\mathrm{ar} = \mathrm{arg}(1 + \delta e^{i \Phi})$.

When $\Phi = 0$, the term $\omega - \omega_+$ becomes negative because the antiresonance frequency is situated between the two photon frequencies, and $\Phi_\mathrm{ar} \rightarrow \Phi_\mathrm{ar} + \pi$. Similar to the case of a system composed of one photon and one magnon, as discussed in \secref{subsec:photon_magnon}, $\Phi_\mathrm{ar} = \pi$. This indicates an opposing phase-jump compared to the phase-jumps of the two resonances.

In the case where $\Phi = \Phi_1 = \pi$ (hence $\Phi_0 = 0$), and $\delta > 1$, the antiresonance has the same phase-jump as the second photon mode with phase factor $\Phi_\mathrm{ar} = \Phi_1 = \pi$. Conversely, when $\delta < 1$, the antiresonance has the same phase-jump as the first photon mode with phase factor $\Phi_\mathrm{ar} = \Phi_0 = 0$. Understanding the phase-jump of the resonances and antiresonance is highly valuable in cavity engineering, as demonstrated in the following sections.

\begin{figure*}[t!]
    \centering
    \includegraphics[width=\textwidth]{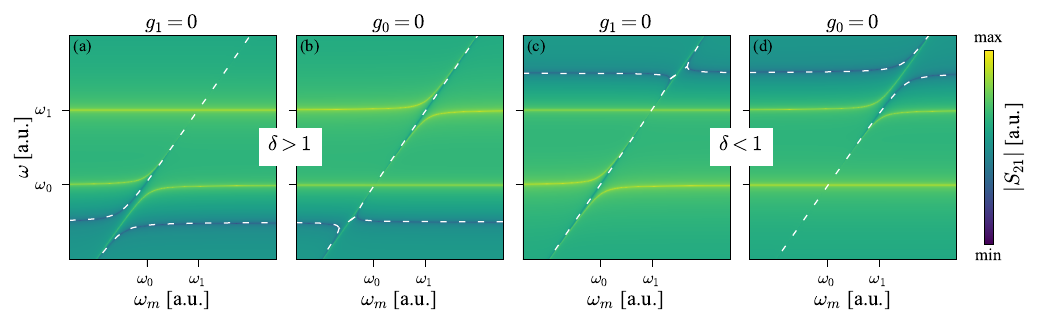}
    \vspace{-1cm}
    \captionsetup{width=\textwidth}
    \caption{Transmission spectra of the system sketched in \figref{fig:systems} (c) considering the 4 different cases related in \tabref{tab:table1} and using \eqref{eq:S_2photon_magnon} showing either an effective level repulsion or level attraction between the antiresonances, where $\delta$ and $\Phi$ are defined in \eqref{eq:antires}, and the effective antiresonance frequency dependences in dashed white lines using \eqref{eq:war}. $\Phi = \pi$ in all four cases.}
    \label{fig:couplings}
\end{figure*}

To conclude on the occurrence of antiresonances, we have explored two simplified scenarios involving only two internal modes. In a real cavity containing an infinity of modes, obtaining an analytical expression for the antiresonance frequency becomes a complex task. Nevertheless, it is feasible to obtain a reasonable approximation of the antiresonance frequency in a real cavity by considering the nearest and most attractive modes numerically, as will be shown later.

\section{Coupling behavior}
\label{sec:coupling}

Here, we clarify all the contributions and conditions required to observe either antiresonance level repulsion or level attraction between two cavity photon modes and one magnon mode. The system is illustrated in \figref{fig:systems} (c).
From \eqref{eq:QLEsol} and \eqref{eq:S}, the transmission of such a system reads:
\begin{widetext}
\begin{equation}
    \label{eq:S_2photon_magnon}
    S_{21} = -i \dfrac{\sqrt{\gamma_{00} \gamma_{01}} (\tilde{\Delta}_1 \Delta_m - g_1^2)  e^{i \Phi_0} + \sqrt{\gamma_{10} \gamma_{11}} (\tilde{\Delta}_0 \Delta_m - g_0^2) e^{i \Phi_1} + \Gamma_4 g_0g_1 - \dfrac{i}{2}\Gamma_2\Delta_m}{\tilde{\Delta}_0 \tilde{\Delta}_1 \Delta_m - g_0^2 \tilde{\Delta}_1 - g_1^2 \tilde{\Delta}_0  + \dfrac{|\Gamma|}{4} \Delta_m + \dfrac{i}{2}(\Gamma + \Gamma^*) g_0 g_1},
\end{equation}
\end{widetext}
where $\Gamma_4 = \sqrt{\gamma_{00} \gamma_{11}} e^{i(\phi_{11} - \phi_{00})} - \sqrt{\gamma_{01} \gamma_{10}}  e^{i(\phi_{10} - \phi_{01})}$.\\
Under the assumption outlined in \eqref{eq:hypothesis}, the system features three resonance frequencies given by:
\begin{equation}
        \Omega_- = \omega_0^-, \qquad
        \Omega_0 = \omega_0^+ + \omega_1^- - \omega_m, \qquad
        \Omega_+ = \omega_1^+,
\end{equation}
where $\omega_0^\pm = \omega_\pm(\tilde{\omega}_0)$, and $\omega_1^\pm = \omega_\pm(\tilde{\omega}_1)$, from \eqref{eq:polaritons}.\\
The minima in transmission of the system are determined by the following antiresonance frequencies:
\begin{equation}
    \label{eq:war}
    \omega_\mathrm{ar}^\pm = \dfrac{1}{2}\left[\omega_\mathrm{ar} + \omega_\mathrm{m} \pm \sqrt{(\omega_\mathrm{ar} - \omega_\mathrm{m})^2 + 4 |g_\mathrm{ar}|^2 e^{i\Phi_\mathrm{ar}}}\right],
\end{equation}
where $\omega_\mathrm{ar}$ is defined in \eqref{eq:antires}. The effective coupling strength between the two antiresonances reads as:
\begin{equation}
    g_\mathrm{ar} = \sqrt{\dfrac{g_1^2 + \delta e^{i\Phi} g_0^2 + C g_0 g_1}{1 + \delta e^{i\Phi}}},
\end{equation}
where $C = \sqrt{\gamma_{11}/\gamma_{01}} e^{i (\phi_{11} - \phi_{01})} + \sqrt{\gamma_{10}/\gamma_{00}} e^{i (\phi_{00} - \phi_{10})}$, $g_0$ and $g_1$ are the coupling strengths of each cavity mode to the magnon, and $\delta$ and $\Phi$ are the same as in \eqref{eq:antires}. When $\Phi = \pi$, $g_\mathrm{ar}$ can be either real or imaginary, depending on the values of $\delta$, $g_0$, $g_1$, and $C$. In \eqref{eq:war}, we chose to explicitly introduce the effective coupling phase $\Phi_\mathrm{ar}$, justifying the absolute value of the effective coupling strength. This choice implies that $\Phi_\mathrm{ar}$ is equal to 0 ($\pi$) when $g_\mathrm{ar}$ is real (imaginary), leading to level repulsion (attraction).

Drawing an analogy with the resonance frequencies of a single photon coupled with a magnon described in \eqref{eq:polaritons}, we can derive the Hamiltonian that governs the antiresonance:
\begin{equation}
    \label{eq:Heff}
    \dfrac{\hat{\mathcal{H}}_\mathrm{ar}}{\hbar} = \omega_\mathrm{ar}\hat{c}^\dagger_\mathrm{ar}\hat{c}_\mathrm{ar} +  \omega_m\hat{m}^\dagger\hat{m} + g_\mathrm{ar} (\hat{c}_\mathrm{ar}^\dagger \hat{m} + e^{i \Phi_\mathrm{ar}} \hat{c}_\mathrm{ar} \hat{m}^\dagger),
\end{equation}
where $\hat{c}_\mathrm{ar}^\dagger$ ($\hat{c}_\mathrm{ar}$) represents the effective creation (annihilation) operator of the cavity antiresonance with an eigenfrequency of $\omega_\mathrm{ar}$. Note that the eigenfrequencies resulting from the diagonalization of this Hamiltonian correspond to the antiresonance frequencies in \eqref{eq:war}.

It is worth noting that while the effective Hamiltonian in \eqref{eq:Heff} and the frequencies of the effective antiresonances in \eqref{eq:war} have been previously introduced based on phenomenological considerations, the effective coupling strength $g_\mathrm{ar}$ lacked a clear physical explanation \cite{raoLevelAttractionLevel2019}. The transmission spectrum of the antiresonance exhibits a level repulsion when the coupling strength $g_\mathrm{ar}$ is real. Conversely, the transmission spectrum shows a level attraction when $g_\mathrm{ar}$ is imaginary.

In the case where only one photonic mode is coupled with the magnon mode and the phase-jump of the two modes are opposed, e.g. their phase factors are $\Phi = \Phi_1 = \pi$ and $\Phi_0 = 0$, the spectrum will exhibit either a repulsive or an attractive signature, and this is determined by the value of $\delta$. The various situations based on the values of $g_0$, $g_1$, and $\delta$, when $\Phi=\pi$ are illustrated in \tabref{tab:table1}.
\begin{table}[b!]
    \caption{Effective coupling behavior}
    \begin{ruledtabular}
        \begin{tabular}{c|cc}
            \label{tab:table1}%
            & $\bm{g_1 = 0}$ & $\bm{g_0 = 0}$ \\
            \hline
            $\bm{\delta > 1}$ & $g_\mathrm{ar} = \dfrac{g_0}{\sqrt{1 - \delta^{-1}}} \in \mathbb{R}$ & $g_\mathrm{ar} = i\dfrac{g_1}{\sqrt{\delta - 1}} \in i\mathbb{R}$ \\
            $\bm{\delta < 1}$ & $g_\mathrm{ar} = i\dfrac{g_0}{\sqrt{\delta^{-1} - 1}} \in i\mathbb{R}$ & $g_\mathrm{ar} = \dfrac{g_1}{\sqrt{1 - \delta}} \in \mathbb{R}$  \\
        \end{tabular}
    \end{ruledtabular}
\end{table}

As mentioned earlier, when $\delta>1$, $\omega_\mathrm{ar} \leq \omega_0$. In this case, the phase-jump of the lower cavity mode and the cavity antiresonance $\hat{c}_\mathrm{ar}$ are opposed. As a result, the lower CMP resulting from the hybridization of the magnon and the lower cavity mode exhibits level repulsion with the antiresonance mode, as depicted in \figref{fig:couplings} (a). However, the the upper cavity mode and the cavity antiresonance exhibit the same phase-jump. Consequently, the lower CMP resulting from the hybridization of the magnon and the upper cavity mode exhibits level attraction with the antiresonance modes, as shown in \figref{fig:couplings} (b).

Conversely, when $\delta$ is less than 1, $\omega_\mathrm{ar} \geq \omega_1$. In this scenario, the phase-jumps ot the upper mode and the cavity antiresonance are opposed, leading to level repulsion at the antiresonance coupling, as depicted in \figref{fig:couplings} (c).
However, the lower cavity mode and the cavity antiresonance exhibit the same phase-jump, leading to level attraction at the antiresonance coupling, as shown in \figref{fig:couplings} (d).

In summary, when there is an opposed phase-jump between a cavity mode and an antiresonance mode, it leads to level repulsion in the coupling between the antiresonance and the CMP resulting from the hybridization of the cavity mode and a magnon. Conversely, when there is a similar phase-jump between a cavity mode and an antiresonance mode, it results in level attraction in the coupling between the antiresonance and the CMP.

As mentioned earlier, in contrast to this two-mode cavity, a real cavity is characterized by an infinite amount of modes coupled to the same magnon mode. This results in a highly complex system, making it challenging to derive a straightforward analytic equation with easy interpretability. Nevertheless, the analysis of the phase of the resonances and the cavity antiresonance proves to be valuable for engineering cavities and predicting the coupling behavior of the antiresonance.

\section{Simulation}

\begin{figure*}[hbt]
    \includegraphics[width=\textwidth]{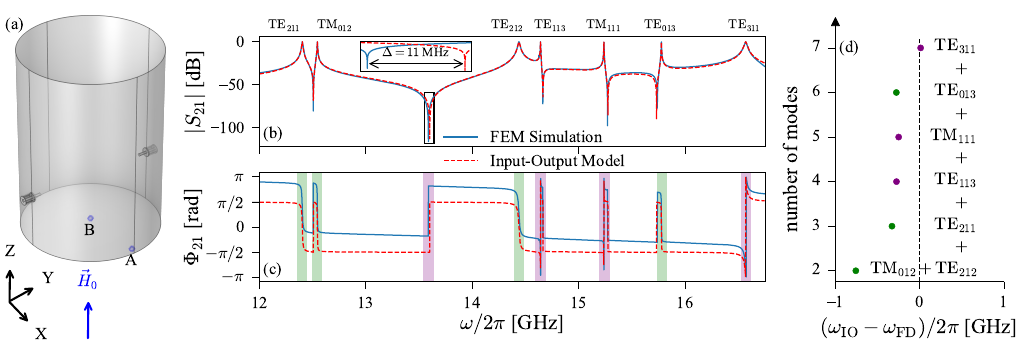}
    \vspace{-1cm}
    \captionsetup{width=\textwidth}
    \caption{(a) Scheme of the cylindrical cavity, whith $A$ the YIG position to observe effective repulsive coupling, and $B$ the YIG position to observe effective attractive coupling; (b) $|S_{21}|$, and (c) $\Phi_{21}$ versus the frequency of the empty cavity, where the input-output model in dashed red line is compared to the FEM simulation in solid blue line. In green areas are the modes exhibiting a negative phase-jump (from $\pi/2$ to $-\pi/2$), and in purple areas are the modes exhibiting a positive phase-jump (from $-\pi/2$ to $\pi/2$), as observed for the phase-jump of the antiresonance at 13.59 GHz in the same color. (d) Graphe illustrating the convergence of the antiresonance frequency between the FEM simulation frequency $\omega_\mathrm{FD}$ and the input-output model frequency $\omega_\mathrm{IO}$ based on the number of considered modes. Colored dots (associated to the added mode) hold the same significance as the colored areas in (c).}
    \label{fig:cavity}
\end{figure*}

\subsection{Model comparison}

This section, focusing on FEM simulation, aims to put into practice the concepts from the previous section applied not only to just two-mode cavity but to a cylindrical cavity exhibiting seven modes in the Ku-band coupled to one magnon mode, represented by a YIG sphere placed at different location.

It has previously been concluded that proximity to a node of the RF $H$-field of a cavity antiresonance was an essential condition for observing the level attraction of an antiresonance \cite{raoLevelAttractionLevel2019}. This effect was attributed to the Lenz effect produced by the cavity, generating a microwave current that hinders the dynamics of the magnetization. It was concluded that the CMP solely repulsively interacted with the antiresonances. Consequently, in an antinode of the RF $H$-field of a cavity antiresonance, the strong repulsive coupling would prevent the manifestation of the Lenz effect.
However, CMP can in fact exert either repulsive or attractive influence on the coupling with the antiresonance. This behavior depends on the phase-jumps of both the antiresonance and the hybridized cavity mode.

A plausible initial approach to modeling a cavity with infinite modes involves considering the minimum number of modes necessary to achieve the same antiresonance frequency. Identifying the phase-jump of an antiresonance provides insights into whether a cavity mode, through the CMP resulting from its coupling with a magnon, behaves repulsively or attractively with the antiresonance.

\subsection{Cavity features}
\label{sec:cavity_features}

\begin{figure*}[htb]
    \centering
    \includegraphics[width=\textwidth]{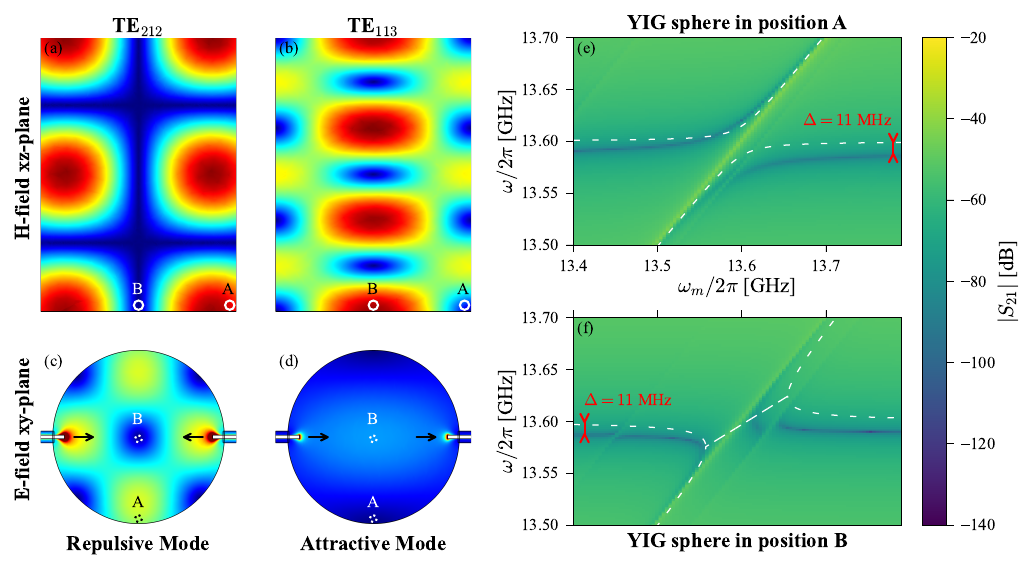}
    \vspace{-1cm}
    \captionsetup{width=\textwidth}
    \caption{Norm of the $H$-field in the $xz$-plane for (a) TE$_{212}$ and (b) TE$_{113}$. Norm of the $E$-field in the $xy$-plane at the height of the probes for (c) TE$_{212}$, and (d) TE$_{113}$. In (a)-(d), the circles illustrate the positions (A or B) of the YIG sphere mentioned in \figref{fig:cavity}, and the black arrows in (c) and (d) indicate  the polarization of the $E$-field at the probe locations. The transmission spectra from FD simulation are shown in (e) when the YIG sphere is placed at position A, and in (f) when the sphere is placed at position B.
    Utilizing the parameter values from \tabref{tab:Rao}, including the mode frequencies, the phase of the $E$-field polarization at each probe, the quality factors, and the coupling strengths of the 7 modes, the white dashed line in (e) and (f) represents the antiresonance frequency obtained by fitting the resulting spectra from the input-ouput model sketched in \figref{fig:input_output_fit} in \appref{sec:input_output_fit}.}
    \label{fig:field_distribution}
\end{figure*}

In this study, a cylindrical cavity described in \citeall{raoLevelAttractionLevel2019} is utilized, featuring a height of 35 mm and a radius of 12.5 mm. Positioned at each side of the cavity are two electrical probes, arranged in parallel as displayed in \figref{fig:cavity} (a). These probes are situated at 10 mm from the cavity's bottom.
A YIG sphere, having a radius of 0.5 mm, is positioned at two distinct locations: position A (x, y, z = 11.9, 0, 0.6 mm from the bottom center of the cavity), inducing level repulsion; and position B (x, y, z = 0, 0, 0.6 mm), inducing level attraction.

In \figref{fig:cavity} (b) and (c), the amplitude and the phase of the transmission are respectively depicted. Seven cavity modes were considered in the input-output model to match the same $|S_{21}|$ trace from the Frequency Domain (FD) simulations. 
To accomplish this, we solve for the eigenmodes of the system first to extract various parameters, including the phases of the $E$-field located at each probes, the quality factor, and the cavity mode frequencies, as detailed in \tabref{tab:Rao}. It is worth noting that our assumption involved equal coupling of a cavity mode with each probe, resulting in equal external coupling strengths between an internal mode and ports.
As depicted in the inset of \figref{fig:cavity} (b), there is a frequency shift of $\Delta = 11$ MHz observed for the antiresonance around 13.59 GHz between the FD simulation and the input-output model, corresponding to the frequencies of 13.589 GHz and 13.600 GHz respectively. 

Note that we need to consider a sufficient number of modes to approach the FD simulation closely as illustrated in \figref{fig:cavity} (d), which shows the dependence of the mismatch between FD simulation and input-output as function of the number of modes considered. The input-output model demonstrates consistent phase, resembling the FEM simulation, except for a constant phase shift proportional to the frequency in the FEM simulation, as shown in \figref{fig:cavity} (c). The absence of this shift in the input-output model has no significant impact on the study. 
Furthermore, with the inclusion of more than 7 modes, the model fails to converge for the antiresonance frequency and cannot accurately match the frequency obtained from FEM simulations.
This phenomenon may be attributed to the first Markov approximation, where the external coupling is assumed to be independent of frequency \cite{xiongExactNonMarkovianCavity2010,moyBornMarkovApproximations1999}. Consequently, there is a significant influence from far-detuned modes on the antiresonance frequency. However, this shifts the antiresonance frequency without altering its phase-jump, which remains the key feature for predicting whether the effective antiresonance coupling would be repulsive or attractive.

Around the antiresonance at a frequency of 13.59 GHz, the phase undergoes a positive phase-jump, from $-\pi/2$ to $\pi/2$. On the contrary, the resonances may exhibit a positive or negative phase-jumps, depicted in purple or green areas respectively in \figref{fig:cavity} (c), and are summarized in \tabref{tab:Rao}. As mentioned earlier, a resonance with the same phase-jump contributes attractively, while a resonance with an opposite phase-jump acts repulsively with the antiresonance.

The TM$_{\theta r z}$ and TE$_{\theta r z}$ modes are characterized by the
number of field nodes in the $\theta$ (spanning 180°), $r$, and $z$ directions, where $r$ represents the radial direction and $\theta$ symbolizes the azimuthal direction in the $xy$-plane.
Referring to \figref{fig:cavity}, for a specified $r$ value (in this case, equal to 1), only the $\theta$ number will determine the coupling phase dependence associated with each mode since the probes are oriented in the radial direction. In this scenario, modes with an odd $\theta$ number induce a repulsive effect, while modes with an even $\theta$ number generate an attractive influence.

\begin{table}[b!]
    \caption{Cavity Modes characteristics}
    \begin{ruledtabular}
        \begin{tabular}{c|cccccc}
            \label{tab:Rao}%
            \multirow{2}{*}{\textbf{Coupling}} & \bf $\bm{\Phi_{21}}$- & \multirow{2}{*}{\textbf{Mode}} & $\bm{\omega/2\pi}$ & \multirow{2}{*}{$\bm{Q}$} & $\bf \bm{g}_{A}/2\bm{\pi}$ & $\bf \bm{g}_{B}/2\bm{\pi}$\\
            & \textbf{jump} && \textbf{[GHz]} && \textbf{[MHz]} & \textbf{[MHz]}\\
            \hline
            \multirow{4}{*}{\textbf{repulsive}}&\multirow{4}{*}{-} & $\mathrm{TE}_{211}$ & 12.4 & 1525 & 13.5 & 0.0\\
            && $\mathrm{TM}_{012}$ & 12.5 & 4441 & 39.3 & 0.0\\
            && $\mathrm{TE}_{212}$ & 14.4 & 912 & 30.1 & 0.0\\
            && $\mathrm{TM}_{013}$ & 15.8 & 9228 & 43.3 & 0.0\\
            \hline
            \multirow{3}{*}{\textbf{attractive}}&\multirow{3}{*}{+} & $\mathrm{TE}_{113}$ & 14.6 & 7501 & 3.6 & 50.0\\
            && $\mathrm{TM}_{111}$ & 15.2 & 12023 & 3.2 & 72.2\\
            && $\mathrm{TE}_{311}$ & 16.6 & 739 & 2.5  & 4.5
        \end{tabular}
    \end{ruledtabular}
\end{table}

The $H$-field and $E$-field distributions of the repulsive mode TE$_{212}$ is illustrated in \figref{fig:field_distribution} (a) and (c), while the field distributions of the attractive mode TE$_{113}$ is depicted in \figref{fig:field_distribution} (b) and (d). For each mode in the Ku-band, we calculated the coupling strengths \cite{bourhillUniversalCharacterizationCavity2020} for two distinct locations of the YIG sphere, A and B, as summarized in \tabref{tab:Rao}. Here, we present only two modes to illustrate the main message of the paper: regarding the field distribution of modes that act repulsively or attractively to the generated antiresonance permit in determining the optimal positioning of the YIG sphere as function of the needs, level attraction or repulsion behavior with the antiresonance. The conclusion of the field distribution of the two illustrated mode also applies for the other repulsive and attractive modes, as shown in \appref{sec:field_distrib}.

At the bottom right side in the $xz$-plane of the cavity (position A), the YIG sphere is positioned at an anti-node of the $H$-field for the repulsive modes and at a node of the $H$-field for the attractive modes. In contrast, at the bottom center position (position B), the YIG sphere is placed at a node of the $H$-field for the repulsive modes and at an anti-node of the $H$-field for the attractive modes.

The modes are either even (for repulsive modes) or odd (for attractive modes) along the azimuthal direction, determining the sign of the probed $E$-field in transmission. This relationship is depicted in \figref{fig:field_distribution}, which shows the $E$-field distribution in the $xy$-plane. In \figref{fig:field_distribution} (e) and (f), the FD simulation of the transmission spectrum versus the magnon frequency and the applied RF frequency reveals the effective level repulsion of the antiresonance when the YIG sphere is positioned at position A, and the effective level attraction of the antiresonance when the YIG sphere is positioned at position B, respectively. Both types of coupling are effectively replicated by the input-output model, depicted in white dashed lines in \figref{fig:field_distribution} (e) and (f). Note that the effective antiresonance frequencies were obtained by fitting the resulting spectra from the input-output model. These frequencies are illustrated in \ref{sec:input_output_fit} in \ref{sec:input_output_fit}.

Despite the similarity in coupling behavior, there exists a notable disparity in the coupling strengths between the simulation and the model. Specifically, the effective repulsive (attractive) coupling strength is of 26 MHz (15 MHz) in the simulation, contrasting with 14 MHz (24 MHz) in the model when the YIG sphere is positioned at location A (B). This discrepancy may arise from the approximations made for the external coupling strength and the neglected internal photonic damping rates. Additionally, limitations in mode truncation may contribute to an inadequate value.
Nevertheless, our findings demonstrate that modes can exhibit either repulsive or attractive effective coupling, and by strategically engineering cavities to spatially separate the modes, it becomes possible to control the coupling behavior when positioning the YIG sphere at different locations.

\section{Conclusion}

This study has provided valuable insights into the intricate interaction between cavity modes and magnons within quasi-closed cavities. Through the incorporation of the input-output formalism enhanced with a crucial phase factor in external couplings, we have not only successfully replicated antiresonance behavior in simulations but also provided explanations for the intriguing phenomena of antiresonances, encompassing both their level repulsion and level attraction, as observed in the transmission spectrum. Furthermore, we provided and demonstrated the physical underpinnings of the effective antiresonance coupling $g_\mathrm{ar}$.
Understanding the phase-jumps of the antiresonance and the different mode families (e.g. characterized by their symmetries), enables the prediction of the behavior of $g_\mathrm{ar}$ regarding the coupling of mode families.

The reconfigured model introduced here carries promising implications for cavity design, offering a versatile tool to tailor these structures for a wide range of applications, from metrology \cite{caoExceptionalMagneticSensitivity2019,zhangExperimentalObservationExceptional2019} to RF devices \cite{hurstNonHermitianPhysicsMagnetic2022,wangNonreciprocalDoublecarrierFrequency2023} and the domain of quantum devices \cite{agarwalGenerationPaircoherent1986,benattiEnvironmentInducedEntanglement2003,kastoryanodissipativePreparationEntanglement2011,reiterScalabledissipativePreparation2016}.

\section{Acknowledgments}
We acknowledge financial support from Brest Métropole for the PhD funding of Guillaume Bourcin. This work is part of the research program supported by the European Union through the European Regional Development Fund (ERDF), by the Ministry of Higher Education and Research, and Brittany region, through the CPER SpaceTech DroneTech, and the ANR projects ICARUS and MagFunc. Jeremy Bourhill is funded by the Australian Research Council Centre of Excellence for Engineered Quantum Systems, CE170100009 and the Centre of Excellence for Dark Matter Particle Physics.

\appendix

\begin{widetext}
\section{Input-Output Theory}
\label{sec:theory}

\subsection{Heisenberg Equation of Motion}
From \eqref{eq:Hamiltonian} and considering the first Markov approximation given in \eqref{eq:Markov}, Heisenberg Equation of Motion (EoM) for the external modes reads \footnote{In the Heisenberg picture, an observable $\hat{d}$ satisfy: $\dot{\hat{d}} = - \dfrac{i}{\hbar} [\hat{d}, \hat{\mathcal{H}}]$}:
\begin{equation}
    \dot{\hat{b}}_{\omega, n}(t) = -i \omega \hat{b}_{\omega, n}(t) + \dfrac{1}{\sqrt{2\pi}}\sum_p{
        \kappa_{pn} \hat{a}_p(t)
    }.
\end{equation}
The solution of this differential equation reads as:
\begin{equation}
    \label{eq:EoMsol}
    \begin{split}
    \hat{b}_{\omega, n}(t) = &\hat{b}_{\omega, n}^\tau e^{-i \omega(t - \tau)}\\
    &+ \dfrac{1}{\sqrt{2\pi}}\sum_p{
            \kappa_{pn} \int\limits_\tau^t{
                \mathrm{d}t' \, \hat{a}_p(t') e^{-i \omega(t - t')}
            }
        },
    \end{split}
\end{equation}
where $\tau$ is a time reference.\\
Subsequently, we define the polychromatic bosonic operator for each port by considering all the bosonic operators of the same port across all frequencies:
\begin{equation}
    \label{eq:polychrome}
    \hat{b}_{n}^\tau(t) = \dfrac{1}{\sqrt{2\pi}} \int_\mathbb{R}{
        \mathrm{d}\omega \, \hat{b}_{\omega, n}^\tau e^{-i \omega(t - \tau)}
    }.
\end{equation}
From this equation, we define the incoming and outgoing wave operators at each port:
\begin{equation}
    \left\{
    \begin{aligned}
        &   \hat{b}_{n}^{\mathrm{in}}(t) = \hat{b}_{n}^{t_0}(t), \quad t_0=-\infty \\
        &   \hat{b}_{n}^{\mathrm{out}}(t) = \hat{b}_{n}^{t_1}(t), \quad t_1=+\infty
    \end{aligned}
    \right..
\end{equation}

\subsection{Input-Output Relation}

Integrating over $\omega$ on both sides, \eqref{eq:EoMsol} becomes \footnote{Note that $\int\limits_{+\infty}^t \mathrm{d}t' \, f(t') = - \int\limits_t^{+\infty} \mathrm{d}t' \, f(t')$, where $f(t')$ is a time-dependent function.}:
\begin{equation}
    \left\{
    \begin{aligned}
        &\dfrac{1}{\sqrt{2\pi}} \int_\mathbb{R} \mathrm{d}\omega \, \hat{b}_{\omega, n}(t) = 
        \dfrac{1}{\sqrt{2\pi}} \int_\mathbb{R} \mathrm{d}\omega \, \hat{b}_{\omega, n}^{t_0} e^{-i \omega(t - t_0)}
        + \dfrac{1}{2\pi}\sum_p{
                \kappa_{pn} \int\limits_{-\infty}^t{
                    \mathrm{d}t' \, \hat{a}_p(t') \int_\mathbb{R} \mathrm{d}\omega \, e^{-i \omega(t - t')}
                }
            } \\
        & \dfrac{1}{\sqrt{2\pi}} \int_\mathbb{R} \mathrm{d}\omega \, \hat{b}_{\omega, n}(t) = 
        \dfrac{1}{\sqrt{2\pi}} \int_\mathbb{R} \mathrm{d}\omega \, \hat{b}_{\omega, n}^{t_1} e^{-i \omega(t - t_1)}
        - \dfrac{1}{2\pi}\sum_p{
                \kappa_{pn} \int\limits_t^{+\infty}{
                    \mathrm{d}t' \, \hat{a}_p(t') \int_\mathbb{R} \mathrm{d}\omega \, e^{-i \omega(t - t')}
                }
            }
    \end{aligned}
    \right..
\end{equation}
where the first term on the right hand side of the first (second) equation is equal to $\hat{b}_n^{in}(t)$ ($\hat{b}_n^{out}(t)$), and the second term is equal to $\dfrac{1}{2} \sum_p \kappa_{pn}$ ($- \dfrac{1}{2} \sum_p \kappa_{pn}$), according to the following properties:
\begin{equation}
    \label{eq:properties}
    \left\{
    \begin{aligned}
        &   \int_\mathbb{R}{
            \mathrm{d}\omega \, e^{-i \omega(t - t')} = 2\pi \delta(t - t')
        }\\
        &    \int\limits_{-\infty}^{t} \mathrm{d}t' \, \hat{a}_p(t') \delta(t - t') = \int\limits_{t}^{+\infty} \mathrm{d}t' \, \hat{a}_p(t') \delta(t - t') = \dfrac{1}{2} \hat{a}_p(t)
    \end{aligned}
    \right..
\end{equation}
This results in the input-output relation:
\begin{equation}
    \label{eq:input-output}
    \hat{b}_{n}^{\mathrm{out}}(t) = \hat{b}_{n}^{\mathrm{in}}(t) + \sum_p{
        \kappa_{pn} \hat{a}_p(t)
    }.
\end{equation}

\subsection{Quantum Langevin Equation}
Assuming $g_{pq}^* = g_{qp}$, the Quantum Langevin Equation (QLE) reads as \footnotemark[1]:
\begin{equation}
    \label{eq:QLE}
        \dot{\hat{a}}_p(t) = -i \omega_p \hat{a}_p(t) -i \sum_{q\neq p}{
            g_{qp} \hat{a}_q(t)
        } 
        - \dfrac{1}{\sqrt{2\pi}} \sum_n{
            \int_\mathbb{R}{
                \mathrm{d}\omega \, \kappa_{pn}^*(\omega) \hat{b}_{\omega n}(t)
            }
        }.
\end{equation}
Substituting the value of $\hat{b}_{\omega, n}$ from \eqref{eq:EoMsol} for $\tau = t_0$ in the QLE gives rise to:
\begin{equation}
    \dot{\hat{a}}_p(t) = -i \omega_p \hat{a}_p(t) -i \sum_{q\neq p}{
        g_{qp} \hat{a}_q(t)
    } 
    - \dfrac{1}{\sqrt{2\pi}} \sum_n
        \kappa_{pn}^* \left[
            \int_\mathbb{R}{
                \mathrm{d}\omega \, \hat{b}_{\omega, n}^{t_0} e^{-i \omega(t - t_0)}
            } 
            + \kappa_{pn} \sum_{q}{
                \int\limits_{t_0}^{t}{
                    \mathrm{d}t' \, \int_\mathbb{R}{
                        \mathrm{d}\omega \, e^{-i \omega(t - t')} \hat{a}_q(t')
                    }
                }
            }
        \right].
\end{equation}
The properties given in \eqref{eq:properties} lead to:
\begin{equation}
    \label{eq:EoM_a}
    \dot{\hat{a}}_p(t) = -i \omega_p \hat{a}_p(t) -i \sum_{q\neq p}{
        g_{qp} \hat{a}_q(t)
    } 
    - \sum_n{
        \kappa_{pn}^* \left(
            \hat{b}_n^{\mathrm{in}}(t) + \sum_q{
                \dfrac{\kappa_{qn}}{2} \hat{a}_q(t)
            }
        \right)
    }.
\end{equation}
Taking the Fourier transform, \eqref{eq:EoM_a} can be expressed as:
\begin{equation}
    \left(
        \omega - \tilde{\omega_p} + \dfrac{i}{2} \sum_n{
            |\kappa_{pn}|^2
        }
    \right) \hat{a}_p(\omega)
    + \sum_{q \neq p}{
        \left(
            \dfrac{i}{2} \sum_n{
                \kappa_{pn}^* \kappa_{qn} - g_{qp}
            }
        \right)
    } \hat{a}_q(\omega) = -i \sum_n{
        \kappa_{pn}^* \hat{b}_n^{\mathrm{in}}(\omega)
    }.
\end{equation}
In matrix form:
\begin{equation}
    \label{eq:QLEsol}
    \left\{
    \begin{aligned}
        & \mathbf{\Omega} \cdot \hat{\mathbf{a}} = -i \mathbf{K}^* \cdot \hat{\mathbf{b}}_\mathrm{in} \\
        & \Omega_{qp} = (\omega - \omega_p)\delta_{qp}+ \dfrac{i}{2} \sum_n(\kappa_{pn}^* \kappa_{qn}) - g_{qp} 
    \end{aligned}
    \right.,
\end{equation}
where $\hat{\mathbf{a}}$ is the vector containing all $\hat{a}_p(\omega)$ operator components, $\hat{\mathbf{b}}_\mathrm{in}$ is the vector containing all $\hat{b}_n^{\mathrm{in}}(\omega)$ operator components, $\mathbf{K}$ is the $p \times m$ matrix with $\kappa_{pm}$ as components, $\Omega_{qp}$ are the components of the $p \times p$ matrix $\mathbf{\Omega}$, and $\delta_{qp}$ is the Kronecker delta.

\subsection{S-parameters}
Substituting the solution of \eqref{eq:QLEsol} into \eqref{eq:input-output} results in:
\begin{equation}
    \hat{\mathbf{b}}_\mathrm{out} = \mathbf{S} \cdot \hat{\mathbf{b}}_\mathrm{in},
\end{equation}
where the $S$-matrix reads as:
\begin{equation}
    \label{eq:S}
    \mathbf{S} = \mathbb{1} - i \mathbf{K}^\mathrm{t} \cdot \mathbf{\Omega}^{-1} \cdot \mathbf{K}^*,
\end{equation}
where $\mathbb{1}$ is the identity matrix.

The $S$-matrix simplifies the computation of reflection and transmission for complex systems with multiple internal modes and ports.
\end{widetext}

\section{Cavity}

\subsection{Model fitting}
\label{sec:input_output_fit}

\begin{figure}[b!]
    \centering
    \includegraphics[width=\linewidth]{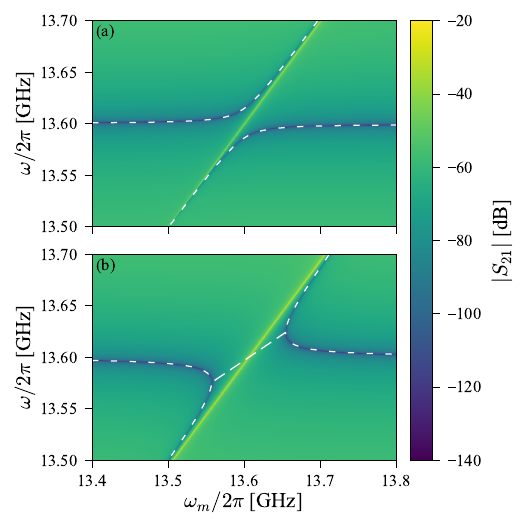}
    \vspace{-1cm}
    \caption{Transmission spectra generated by the input-output model for two different positions of the YIG sphere: (a) when positioned at A, and (b) when positioned at B. The parameter values injected into the model, including mode frequencies, the phase of the $E$-field polarization at each probe, quality factors, and coupling strengths of the 7 modes, are detailed in \tabref{tab:Rao}. The antiresonance frequencies, indicated to the dashed black lines, were determined by fitting the spectra.}
    \label{fig:input_output_fit}
\end{figure}

In Fig. \ref{fig:input_output_fit}, the transmission spectra generated by the input-output model for two distinct positions of the YIG sphere (A and B) are illustrated in (a) and (b) respectively. To generate these spectra, the model was provided with necessary parameter values, including the mode frequencies, the phase of the $E$-field polarization at each probe, the quality factors, and the coupling strengths of the 7 modes as detailed in \tabref{tab:Rao}. Due to the complexity of obtaining an analytical solution for the polariton frequencies, the antiresonance frequencies were determined by fitting the two spectra.

\subsection{Field Distribution of cavity modes}
\label{sec:field_distrib}

\begin{figure*}[ht!]
    \centering
    \includegraphics[width=\textwidth]{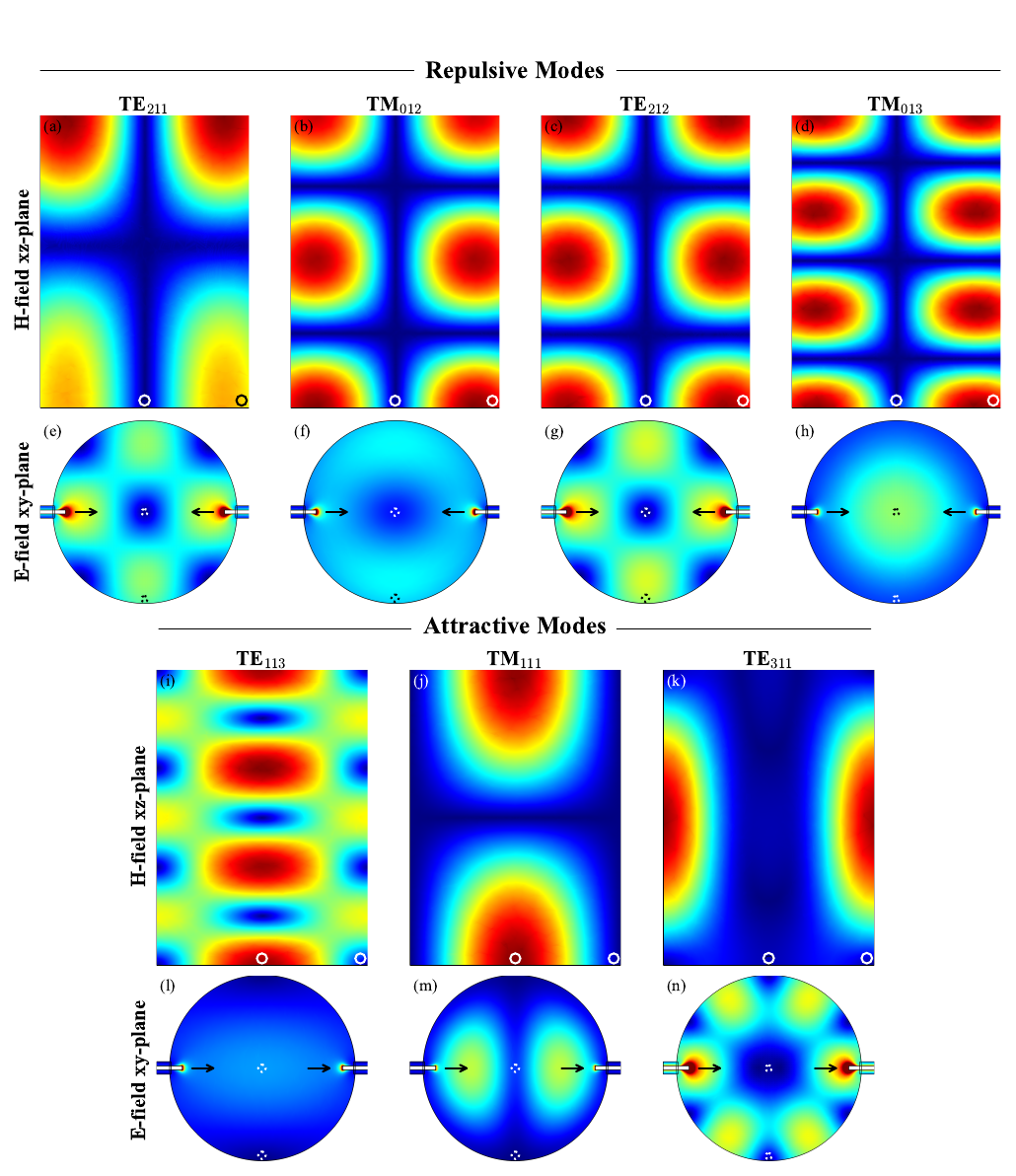}
    \vspace{-1cm}
    \captionsetup{width=\textwidth}
    \caption{Norm of the $H$-field in the $xz$-plane for (a)-(d) repulsive modes and (i)-(k) attractive modes. Norm of the $E$-field in the $xy$-plane at the height of the probes for (e)-(h) repulsive modes, and (l)-(n) attractive modes. The circles illustrate the positions (A or B) of the YIG sphere mentioned in \figref{fig:cavity}, and the black arrows in (e)-(h) and (l)-(n) indicate  the polarization of the $E$-field at the probe locations.}
    \label{fig:all_field_fistribution}
\end{figure*}

\figref{fig:all_field_fistribution} illustrates the magnitudes of the $H$- and $E$-fields for the seven considered cavity modes in the input-output model, as discussed in \secref{sec:cavity_features}.

For $r=1$, the repulsive modes exhibit an even $\theta$ number, leading to a minimum in the $H$-field at position B. Additionally, the $E$-field polarization at the two probes is $\pi$-dephased, a condition necessary for observing level repulsion in the effective coupling of the antiresonance at 13.59 GHz with the coupled magnon-photon mode.

Conversely, with $r=1$, the attractive modes exhibit an odd $\theta$ number, resulting in a minimum of the $H$-field at position A. Furthermore, the $E$-field polarization at the two probes is in phase, a condition for observing level attraction in the effective coupling of the antiresonance at 13.59 GHz with the coupled magnon-photon mode.

\newpage

\bibliography{C:/Users/Guillaume/Documents/Bibliography}

\begin{thebibliography}{42}%
\makeatletter
\providecommand \@ifxundefined [1]{%
 \@ifx{#1\undefined}
}%
\providecommand \@ifnum [1]{%
 \ifnum #1\expandafter \@firstoftwo
 \else \expandafter \@secondoftwo
 \fi
}%
\providecommand \@ifx [1]{%
 \ifx #1\expandafter \@firstoftwo
 \else \expandafter \@secondoftwo
 \fi
}%
\providecommand \natexlab [1]{#1}%
\providecommand \enquote  [1]{``#1''}%
\providecommand \bibnamefont  [1]{#1}%
\providecommand \bibfnamefont [1]{#1}%
\providecommand \citenamefont [1]{#1}%
\providecommand \href@noop [0]{\@secondoftwo}%
\providecommand \href [0]{\begingroup \@sanitize@url \@href}%
\providecommand \@href[1]{\@@startlink{#1}\@@href}%
\providecommand \@@href[1]{\endgroup#1\@@endlink}%
\providecommand \@sanitize@url [0]{\catcode `\\12\catcode `\$12\catcode
  `\&12\catcode `\#12\catcode `\^12\catcode `\_12\catcode `\%12\relax}%
\providecommand \@@startlink[1]{}%
\providecommand \@@endlink[0]{}%
\providecommand \url  [0]{\begingroup\@sanitize@url \@url }%
\providecommand \@url [1]{\endgroup\@href {#1}{\urlprefix }}%
\providecommand \urlprefix  [0]{URL }%
\providecommand \Eprint [0]{\href }%
\providecommand \doibase [0]{https://doi.org/}%
\providecommand \selectlanguage [0]{\@gobble}%
\providecommand \bibinfo  [0]{\@secondoftwo}%
\providecommand \bibfield  [0]{\@secondoftwo}%
\providecommand \translation [1]{[#1]}%
\providecommand \BibitemOpen [0]{}%
\providecommand \bibitemStop [0]{}%
\providecommand \bibitemNoStop [0]{.\EOS\space}%
\providecommand \EOS [0]{\spacefactor3000\relax}%
\providecommand \BibitemShut  [1]{\csname bibitem#1\endcsname}%
\let\auto@bib@innerbib\@empty
\bibitem [{\citenamefont {Barman}\ \emph {et~al.}(2021)\citenamefont {Barman},
  \citenamefont {Gubbiotti}, \citenamefont {Ladak}, \citenamefont {Adeyeye},
  \citenamefont {Krawczyk}, \citenamefont {Gr{\"a}fe}, \citenamefont
  {Adelmann}, \citenamefont {Cotofana}, \citenamefont {Naeemi}, \citenamefont
  {Vasyuchka}, \citenamefont {Hillebrands}, \citenamefont {Nikitov},
  \citenamefont {Yu}, \citenamefont {Grundler}, \citenamefont {Sadovnikov},
  \citenamefont {Grachev}, \citenamefont {Sheshukova}, \citenamefont
  {Duquesne}, \citenamefont {Marangolo}, \citenamefont {Csaba}, \citenamefont
  {Porod}, \citenamefont {Demidov}, \citenamefont {Urazhdin}, \citenamefont
  {Demokritov}, \citenamefont {Albisetti}, \citenamefont {Petti}, \citenamefont
  {Bertacco}, \citenamefont {Schultheiss}, \citenamefont {Kruglyak},
  \citenamefont {Poimanov}, \citenamefont {Sahoo}, \citenamefont {Sinha},
  \citenamefont {Yang}, \citenamefont {M{\"u}nzenberg}, \citenamefont
  {Moriyama}, \citenamefont {Mizukami}, \citenamefont {Landeros}, \citenamefont
  {Gallardo}, \citenamefont {Carlotti}, \citenamefont {Kim}, \citenamefont
  {Stamps}, \citenamefont {Camley}, \citenamefont {Rana}, \citenamefont
  {Otani}, \citenamefont {Yu}, \citenamefont {Yu}, \citenamefont {Bauer},
  \citenamefont {Back}, \citenamefont {Uhrig}, \citenamefont {Dobrovolskiy},
  \citenamefont {Budinska}, \citenamefont {Qin}, \citenamefont {{van Dijken}},
  \citenamefont {Chumak}, \citenamefont {Khitun}, \citenamefont {Nikonov},
  \citenamefont {Young}, \citenamefont {Zingsem},\ and\ \citenamefont
  {Winklhofer}}]{barman2021MagnonicsRoadmap2021}%
  \BibitemOpen
  \bibfield  {author} {\bibinfo {author} {\bibfnamefont {A.}~\bibnamefont
  {Barman}}, \bibinfo {author} {\bibfnamefont {G.}~\bibnamefont {Gubbiotti}},
  \bibinfo {author} {\bibfnamefont {S.}~\bibnamefont {Ladak}}, \bibinfo
  {author} {\bibfnamefont {A.~O.}\ \bibnamefont {Adeyeye}}, \bibinfo {author}
  {\bibfnamefont {M.}~\bibnamefont {Krawczyk}}, \bibinfo {author}
  {\bibfnamefont {J.}~\bibnamefont {Gr{\"a}fe}}, \bibinfo {author}
  {\bibfnamefont {C.}~\bibnamefont {Adelmann}}, \bibinfo {author}
  {\bibfnamefont {S.}~\bibnamefont {Cotofana}}, \bibinfo {author}
  {\bibfnamefont {A.}~\bibnamefont {Naeemi}}, \bibinfo {author} {\bibfnamefont
  {V.~I.}\ \bibnamefont {Vasyuchka}}, \bibinfo {author} {\bibfnamefont
  {B.}~\bibnamefont {Hillebrands}}, \bibinfo {author} {\bibfnamefont {S.~A.}\
  \bibnamefont {Nikitov}}, \bibinfo {author} {\bibfnamefont {H.}~\bibnamefont
  {Yu}}, \bibinfo {author} {\bibfnamefont {D.}~\bibnamefont {Grundler}},
  \bibinfo {author} {\bibfnamefont {A.~V.}\ \bibnamefont {Sadovnikov}},
  \bibinfo {author} {\bibfnamefont {A.~A.}\ \bibnamefont {Grachev}}, \bibinfo
  {author} {\bibfnamefont {S.~E.}\ \bibnamefont {Sheshukova}}, \bibinfo
  {author} {\bibfnamefont {J.-Y.}\ \bibnamefont {Duquesne}}, \bibinfo {author}
  {\bibfnamefont {M.}~\bibnamefont {Marangolo}}, \bibinfo {author}
  {\bibfnamefont {G.}~\bibnamefont {Csaba}}, \bibinfo {author} {\bibfnamefont
  {W.}~\bibnamefont {Porod}}, \bibinfo {author} {\bibfnamefont {V.~E.}\
  \bibnamefont {Demidov}}, \bibinfo {author} {\bibfnamefont {S.}~\bibnamefont
  {Urazhdin}}, \bibinfo {author} {\bibfnamefont {S.~O.}\ \bibnamefont
  {Demokritov}}, \bibinfo {author} {\bibfnamefont {E.}~\bibnamefont
  {Albisetti}}, \bibinfo {author} {\bibfnamefont {D.}~\bibnamefont {Petti}},
  \bibinfo {author} {\bibfnamefont {R.}~\bibnamefont {Bertacco}}, \bibinfo
  {author} {\bibfnamefont {H.}~\bibnamefont {Schultheiss}}, \bibinfo {author}
  {\bibfnamefont {V.~V.}\ \bibnamefont {Kruglyak}}, \bibinfo {author}
  {\bibfnamefont {V.~D.}\ \bibnamefont {Poimanov}}, \bibinfo {author}
  {\bibfnamefont {S.}~\bibnamefont {Sahoo}}, \bibinfo {author} {\bibfnamefont
  {J.}~\bibnamefont {Sinha}}, \bibinfo {author} {\bibfnamefont
  {H.}~\bibnamefont {Yang}}, \bibinfo {author} {\bibfnamefont {M.}~\bibnamefont
  {M{\"u}nzenberg}}, \bibinfo {author} {\bibfnamefont {T.}~\bibnamefont
  {Moriyama}}, \bibinfo {author} {\bibfnamefont {S.}~\bibnamefont {Mizukami}},
  \bibinfo {author} {\bibfnamefont {P.}~\bibnamefont {Landeros}}, \bibinfo
  {author} {\bibfnamefont {R.~A.}\ \bibnamefont {Gallardo}}, \bibinfo {author}
  {\bibfnamefont {G.}~\bibnamefont {Carlotti}}, \bibinfo {author}
  {\bibfnamefont {J.-V.}\ \bibnamefont {Kim}}, \bibinfo {author} {\bibfnamefont
  {R.~L.}\ \bibnamefont {Stamps}}, \bibinfo {author} {\bibfnamefont {R.~E.}\
  \bibnamefont {Camley}}, \bibinfo {author} {\bibfnamefont {B.}~\bibnamefont
  {Rana}}, \bibinfo {author} {\bibfnamefont {Y.}~\bibnamefont {Otani}},
  \bibinfo {author} {\bibfnamefont {W.}~\bibnamefont {Yu}}, \bibinfo {author}
  {\bibfnamefont {T.}~\bibnamefont {Yu}}, \bibinfo {author} {\bibfnamefont
  {G.~E.~W.}\ \bibnamefont {Bauer}}, \bibinfo {author} {\bibfnamefont
  {C.}~\bibnamefont {Back}}, \bibinfo {author} {\bibfnamefont {G.~S.}\
  \bibnamefont {Uhrig}}, \bibinfo {author} {\bibfnamefont {O.~V.}\ \bibnamefont
  {Dobrovolskiy}}, \bibinfo {author} {\bibfnamefont {B.}~\bibnamefont
  {Budinska}}, \bibinfo {author} {\bibfnamefont {H.}~\bibnamefont {Qin}},
  \bibinfo {author} {\bibfnamefont {S.}~\bibnamefont {{van Dijken}}}, \bibinfo
  {author} {\bibfnamefont {A.~V.}\ \bibnamefont {Chumak}}, \bibinfo {author}
  {\bibfnamefont {A.}~\bibnamefont {Khitun}}, \bibinfo {author} {\bibfnamefont
  {D.~E.}\ \bibnamefont {Nikonov}}, \bibinfo {author} {\bibfnamefont {I.~A.}\
  \bibnamefont {Young}}, \bibinfo {author} {\bibfnamefont {B.~W.}\ \bibnamefont
  {Zingsem}},\ and\ \bibinfo {author} {\bibfnamefont {M.}~\bibnamefont
  {Winklhofer}},\ }\bibfield  {title} {\bibinfo {title} {The 2021 {{Magnonics
  Roadmap}}},\ }\href {https://doi.org/10.1088/1361-648X/abec1a} {\bibfield
  {journal} {\bibinfo  {journal} {Journal of Physics: Condensed Matter}\
  }\textbf {\bibinfo {volume} {33}},\ \bibinfo {pages} {413001} (\bibinfo
  {year} {2021})}\BibitemShut {NoStop}%
\bibitem [{\citenamefont {{Lachance-Quirion}}\ \emph
  {et~al.}(2019)\citenamefont {{Lachance-Quirion}}, \citenamefont {Tabuchi},
  \citenamefont {Gloppe}, \citenamefont {Usami},\ and\ \citenamefont
  {Nakamura}}]{lachance-quirionHybridQuantumSystems2019}%
  \BibitemOpen
  \bibfield  {author} {\bibinfo {author} {\bibfnamefont {D.}~\bibnamefont
  {{Lachance-Quirion}}}, \bibinfo {author} {\bibfnamefont {Y.}~\bibnamefont
  {Tabuchi}}, \bibinfo {author} {\bibfnamefont {A.}~\bibnamefont {Gloppe}},
  \bibinfo {author} {\bibfnamefont {K.}~\bibnamefont {Usami}},\ and\ \bibinfo
  {author} {\bibfnamefont {Y.}~\bibnamefont {Nakamura}},\ }\bibfield  {title}
  {\bibinfo {title} {Hybrid quantum systems based on magnonics},\ }\href
  {https://doi.org/10.7567/1882-0786/ab248d} {\bibfield  {journal} {\bibinfo
  {journal} {Applied Physics Express}\ }\textbf {\bibinfo {volume} {12}},\
  \bibinfo {pages} {070101} (\bibinfo {year} {2019})}\BibitemShut {NoStop}%
\bibitem [{\citenamefont {Zare~Rameshti}\ \emph {et~al.}(2022)\citenamefont
  {Zare~Rameshti}, \citenamefont {Viola~Kusminskiy}, \citenamefont {Haigh},
  \citenamefont {Usami}, \citenamefont {{Lachance-Quirion}}, \citenamefont
  {Nakamura}, \citenamefont {Hu}, \citenamefont {Tang}, \citenamefont {Bauer},\
  and\ \citenamefont {Blanter}}]{zarerameshtiCavityMagnonics2022}%
  \BibitemOpen
  \bibfield  {author} {\bibinfo {author} {\bibfnamefont {B.}~\bibnamefont
  {Zare~Rameshti}}, \bibinfo {author} {\bibfnamefont {S.}~\bibnamefont
  {Viola~Kusminskiy}}, \bibinfo {author} {\bibfnamefont {J.~A.}\ \bibnamefont
  {Haigh}}, \bibinfo {author} {\bibfnamefont {K.}~\bibnamefont {Usami}},
  \bibinfo {author} {\bibfnamefont {D.}~\bibnamefont {{Lachance-Quirion}}},
  \bibinfo {author} {\bibfnamefont {Y.}~\bibnamefont {Nakamura}}, \bibinfo
  {author} {\bibfnamefont {C.-M.}\ \bibnamefont {Hu}}, \bibinfo {author}
  {\bibfnamefont {H.~X.}\ \bibnamefont {Tang}}, \bibinfo {author}
  {\bibfnamefont {G.~E.}\ \bibnamefont {Bauer}},\ and\ \bibinfo {author}
  {\bibfnamefont {Y.~M.}\ \bibnamefont {Blanter}},\ }\bibfield  {title}
  {\bibinfo {title} {Cavity magnonics},\ }\href
  {https://doi.org/10.1016/j.physrep.2022.06.001} {\bibfield  {journal}
  {\bibinfo  {journal} {Physics Reports}\ }\textbf {\bibinfo {volume} {979}},\
  \bibinfo {pages} {1} (\bibinfo {year} {2022})}\BibitemShut {NoStop}%
\bibitem [{\citenamefont {Harder}\ \emph {et~al.}(2021)\citenamefont {Harder},
  \citenamefont {Yao}, \citenamefont {Gui},\ and\ \citenamefont
  {Hu}}]{hardercoherentdissipativeCavity2021}%
  \BibitemOpen
  \bibfield  {author} {\bibinfo {author} {\bibfnamefont {M.}~\bibnamefont
  {Harder}}, \bibinfo {author} {\bibfnamefont {B.~M.}\ \bibnamefont {Yao}},
  \bibinfo {author} {\bibfnamefont {Y.~S.}\ \bibnamefont {Gui}},\ and\ \bibinfo
  {author} {\bibfnamefont {C.-M.}\ \bibnamefont {Hu}},\ }\bibfield  {title}
  {\bibinfo {title} {Coherent and dissipative cavity magnonics},\ }\href
  {https://doi.org/10.1063/5.0046202} {\bibfield  {journal} {\bibinfo
  {journal} {Journal of Applied Physics}\ }\textbf {\bibinfo {volume} {129}},\
  \bibinfo {pages} {201101} (\bibinfo {year} {2021})}\BibitemShut {NoStop}%
\bibitem [{\citenamefont {Wang}\ and\ \citenamefont
  {Hu}(2020)}]{wangdissipativeCouplingsCavity2020}%
  \BibitemOpen
  \bibfield  {author} {\bibinfo {author} {\bibfnamefont {Y.-P.}\ \bibnamefont
  {Wang}}\ and\ \bibinfo {author} {\bibfnamefont {C.-M.}\ \bibnamefont {Hu}},\
  }\bibfield  {title} {\bibinfo {title} {Dissipative couplings in cavity
  magnonics},\ }\href {https://doi.org/10.1063/1.5144202} {\bibfield  {journal}
  {\bibinfo  {journal} {Journal of Applied Physics}\ }\textbf {\bibinfo
  {volume} {127}},\ \bibinfo {pages} {130901} (\bibinfo {year}
  {2020})}\BibitemShut {NoStop}%
\bibitem [{\citenamefont {Harder}\ \emph {et~al.}(2018)\citenamefont {Harder},
  \citenamefont {Yang}, \citenamefont {Yao}, \citenamefont {Yu}, \citenamefont
  {Rao}, \citenamefont {Gui}, \citenamefont {Stamps},\ and\ \citenamefont
  {Hu}}]{harderLevelAttractionDue2018}%
  \BibitemOpen
  \bibfield  {author} {\bibinfo {author} {\bibfnamefont {M.}~\bibnamefont
  {Harder}}, \bibinfo {author} {\bibfnamefont {Y.}~\bibnamefont {Yang}},
  \bibinfo {author} {\bibfnamefont {B.~M.}\ \bibnamefont {Yao}}, \bibinfo
  {author} {\bibfnamefont {C.~H.}\ \bibnamefont {Yu}}, \bibinfo {author}
  {\bibfnamefont {J.~W.}\ \bibnamefont {Rao}}, \bibinfo {author} {\bibfnamefont
  {Y.~S.}\ \bibnamefont {Gui}}, \bibinfo {author} {\bibfnamefont {R.~L.}\
  \bibnamefont {Stamps}},\ and\ \bibinfo {author} {\bibfnamefont {C.-M.}\
  \bibnamefont {Hu}},\ }\bibfield  {title} {\bibinfo {title} {Level
  {{Attraction Due}} to {{Dissipative Magnon-Photon Coupling}}},\ }\href
  {https://doi.org/10.1103/PhysRevLett.121.137203} {\bibfield  {journal}
  {\bibinfo  {journal} {Physical Review Letters}\ }\textbf {\bibinfo {volume}
  {121}},\ \bibinfo {pages} {137203} (\bibinfo {year} {2018})}\BibitemShut
  {NoStop}%
\bibitem [{\citenamefont {Bhoi}\ \emph {et~al.}(2019)\citenamefont {Bhoi},
  \citenamefont {Kim}, \citenamefont {Jang}, \citenamefont {Kim}, \citenamefont
  {Yang}, \citenamefont {Cho},\ and\ \citenamefont
  {Kim}}]{bhoiAbnormalAnticrossingEffect2019}%
  \BibitemOpen
  \bibfield  {author} {\bibinfo {author} {\bibfnamefont {B.}~\bibnamefont
  {Bhoi}}, \bibinfo {author} {\bibfnamefont {B.}~\bibnamefont {Kim}}, \bibinfo
  {author} {\bibfnamefont {S.-H.}\ \bibnamefont {Jang}}, \bibinfo {author}
  {\bibfnamefont {J.}~\bibnamefont {Kim}}, \bibinfo {author} {\bibfnamefont
  {J.}~\bibnamefont {Yang}}, \bibinfo {author} {\bibfnamefont {Y.-J.}\
  \bibnamefont {Cho}},\ and\ \bibinfo {author} {\bibfnamefont {S.-K.}\
  \bibnamefont {Kim}},\ }\bibfield  {title} {\bibinfo {title} {Abnormal
  anticrossing effect in photon-magnon coupling},\ }\href
  {https://doi.org/10.1103/PhysRevB.99.134426} {\bibfield  {journal} {\bibinfo
  {journal} {Physical Review B}\ }\textbf {\bibinfo {volume} {99}},\ \bibinfo
  {pages} {134426} (\bibinfo {year} {2019})}\BibitemShut {NoStop}%
\bibitem [{\citenamefont {Rao}\ \emph {et~al.}(2020)\citenamefont {Rao},
  \citenamefont {Wang}, \citenamefont {Yang}, \citenamefont {Yu}, \citenamefont
  {Gui}, \citenamefont {Fan}, \citenamefont {Xue},\ and\ \citenamefont
  {Hu}}]{raoInteractionsMagnonMode2020}%
  \BibitemOpen
  \bibfield  {author} {\bibinfo {author} {\bibfnamefont {J.~W.}\ \bibnamefont
  {Rao}}, \bibinfo {author} {\bibfnamefont {Y.~P.}\ \bibnamefont {Wang}},
  \bibinfo {author} {\bibfnamefont {Y.}~\bibnamefont {Yang}}, \bibinfo {author}
  {\bibfnamefont {T.}~\bibnamefont {Yu}}, \bibinfo {author} {\bibfnamefont
  {Y.~S.}\ \bibnamefont {Gui}}, \bibinfo {author} {\bibfnamefont {X.~L.}\
  \bibnamefont {Fan}}, \bibinfo {author} {\bibfnamefont {D.~S.}\ \bibnamefont
  {Xue}},\ and\ \bibinfo {author} {\bibfnamefont {C.-M.}\ \bibnamefont {Hu}},\
  }\bibfield  {title} {\bibinfo {title} {Interactions between a magnon mode and
  a cavity photon mode mediated by traveling photons},\ }\href
  {https://doi.org/10.1103/PhysRevB.101.064404} {\bibfield  {journal} {\bibinfo
   {journal} {Physical Review B}\ }\textbf {\bibinfo {volume} {101}},\ \bibinfo
  {pages} {064404} (\bibinfo {year} {2020})}\BibitemShut {NoStop}%
\bibitem [{\citenamefont {Yao}\ \emph {et~al.}(2019{\natexlab{a}})\citenamefont
  {Yao}, \citenamefont {Yu}, \citenamefont {Gui}, \citenamefont {Rao},
  \citenamefont {Zhao}, \citenamefont {Lu},\ and\ \citenamefont
  {Hu}}]{yaocoherentControlMagnon2019}%
  \BibitemOpen
  \bibfield  {author} {\bibinfo {author} {\bibfnamefont {B.}~\bibnamefont
  {Yao}}, \bibinfo {author} {\bibfnamefont {T.}~\bibnamefont {Yu}}, \bibinfo
  {author} {\bibfnamefont {Y.~S.}\ \bibnamefont {Gui}}, \bibinfo {author}
  {\bibfnamefont {J.~W.}\ \bibnamefont {Rao}}, \bibinfo {author} {\bibfnamefont
  {Y.~T.}\ \bibnamefont {Zhao}}, \bibinfo {author} {\bibfnamefont
  {W.}~\bibnamefont {Lu}},\ and\ \bibinfo {author} {\bibfnamefont {C.-M.}\
  \bibnamefont {Hu}},\ }\bibfield  {title} {\bibinfo {title} {Coherent control
  of magnon radiative damping with local photon states},\ }\href
  {https://doi.org/10.1038/s42005-019-0264-z} {\bibfield  {journal} {\bibinfo
  {journal} {Communications Physics}\ }\textbf {\bibinfo {volume} {2}},\
  \bibinfo {pages} {1} (\bibinfo {year} {2019}{\natexlab{a}})}\BibitemShut
  {NoStop}%
\bibitem [{\citenamefont {Yao}\ \emph {et~al.}(2019{\natexlab{b}})\citenamefont
  {Yao}, \citenamefont {Yu}, \citenamefont {Zhang}, \citenamefont {Lu},
  \citenamefont {Gui}, \citenamefont {Hu},\ and\ \citenamefont
  {Blanter}}]{yaoMicroscopicOriginMagnonphoton2019}%
  \BibitemOpen
  \bibfield  {author} {\bibinfo {author} {\bibfnamefont {B.}~\bibnamefont
  {Yao}}, \bibinfo {author} {\bibfnamefont {T.}~\bibnamefont {Yu}}, \bibinfo
  {author} {\bibfnamefont {X.}~\bibnamefont {Zhang}}, \bibinfo {author}
  {\bibfnamefont {W.}~\bibnamefont {Lu}}, \bibinfo {author} {\bibfnamefont
  {Y.}~\bibnamefont {Gui}}, \bibinfo {author} {\bibfnamefont {C.-M.}\
  \bibnamefont {Hu}},\ and\ \bibinfo {author} {\bibfnamefont {Y.~M.}\
  \bibnamefont {Blanter}},\ }\bibfield  {title} {\bibinfo {title} {The
  microscopic origin of magnon-photon level attraction by traveling waves:
  {{Theory}} and experiment},\ }\href
  {https://doi.org/10.1103/PhysRevB.100.214426} {\bibfield  {journal} {\bibinfo
   {journal} {Physical Review B}\ }\textbf {\bibinfo {volume} {100}},\ \bibinfo
  {pages} {214426} (\bibinfo {year} {2019}{\natexlab{b}})}\BibitemShut
  {NoStop}%
\bibitem [{\citenamefont {Yang}\ \emph {et~al.}(2019)\citenamefont {Yang},
  \citenamefont {Rao}, \citenamefont {Gui}, \citenamefont {Yao}, \citenamefont
  {Lu},\ and\ \citenamefont {Hu}}]{yangControlMagnonphotonLevel2019}%
  \BibitemOpen
  \bibfield  {author} {\bibinfo {author} {\bibfnamefont {Y.}~\bibnamefont
  {Yang}}, \bibinfo {author} {\bibfnamefont {J.}~\bibnamefont {Rao}}, \bibinfo
  {author} {\bibfnamefont {Y.}~\bibnamefont {Gui}}, \bibinfo {author}
  {\bibfnamefont {B.}~\bibnamefont {Yao}}, \bibinfo {author} {\bibfnamefont
  {W.}~\bibnamefont {Lu}},\ and\ \bibinfo {author} {\bibfnamefont {C.-M.}\
  \bibnamefont {Hu}},\ }\bibfield  {title} {\bibinfo {title} {Control of the
  {{Magnon-Photon Level Attraction}} in a {{Planar Cavity}}},\ }\href
  {https://doi.org/10.1103/PhysRevApplied.11.054023} {\bibfield  {journal}
  {\bibinfo  {journal} {Physical Review Applied}\ }\textbf {\bibinfo {volume}
  {11}},\ \bibinfo {pages} {054023} (\bibinfo {year} {2019})}\BibitemShut
  {NoStop}%
\bibitem [{\citenamefont {Rao}\ \emph {et~al.}(2019)\citenamefont {Rao},
  \citenamefont {Yu}, \citenamefont {Zhao}, \citenamefont {Gui}, \citenamefont
  {Fan}, \citenamefont {Xue},\ and\ \citenamefont
  {Hu}}]{raoLevelAttractionLevel2019}%
  \BibitemOpen
  \bibfield  {author} {\bibinfo {author} {\bibfnamefont {J.~W.}\ \bibnamefont
  {Rao}}, \bibinfo {author} {\bibfnamefont {C.~H.}\ \bibnamefont {Yu}},
  \bibinfo {author} {\bibfnamefont {Y.~T.}\ \bibnamefont {Zhao}}, \bibinfo
  {author} {\bibfnamefont {Y.~S.}\ \bibnamefont {Gui}}, \bibinfo {author}
  {\bibfnamefont {X.~L.}\ \bibnamefont {Fan}}, \bibinfo {author} {\bibfnamefont
  {D.~S.}\ \bibnamefont {Xue}},\ and\ \bibinfo {author} {\bibfnamefont {C.-M.}\
  \bibnamefont {Hu}},\ }\bibfield  {title} {\bibinfo {title} {Level attraction
  and level repulsion of magnon coupled with a cavity anti-resonance},\ }\href
  {https://doi.org/10.1088/1367-2630/ab2482} {\bibfield  {journal} {\bibinfo
  {journal} {New Journal of Physics}\ }\textbf {\bibinfo {volume} {21}},\
  \bibinfo {pages} {065001} (\bibinfo {year} {2019})}\BibitemShut {NoStop}%
\bibitem [{\citenamefont {Grigoryan}\ and\ \citenamefont
  {Xia}(2019)}]{grigoryanCavitymediateddissipativeSpinspin2019}%
  \BibitemOpen
  \bibfield  {author} {\bibinfo {author} {\bibfnamefont {V.~L.}\ \bibnamefont
  {Grigoryan}}\ and\ \bibinfo {author} {\bibfnamefont {K.}~\bibnamefont
  {Xia}},\ }\bibfield  {title} {\bibinfo {title} {Cavity-mediated dissipative
  spin-spin coupling},\ }\href {https://doi.org/10.1103/PhysRevB.100.014415}
  {\bibfield  {journal} {\bibinfo  {journal} {Physical Review B}\ }\textbf
  {\bibinfo {volume} {100}},\ \bibinfo {pages} {014415} (\bibinfo {year}
  {2019})}\BibitemShut {NoStop}%
\bibitem [{\citenamefont {Reiter}\ \emph {et~al.}(2016)\citenamefont {Reiter},
  \citenamefont {Reeb},\ and\ \citenamefont
  {S{\o}rensen}}]{reiterScalabledissipativePreparation2016}%
  \BibitemOpen
  \bibfield  {author} {\bibinfo {author} {\bibfnamefont {F.}~\bibnamefont
  {Reiter}}, \bibinfo {author} {\bibfnamefont {D.}~\bibnamefont {Reeb}},\ and\
  \bibinfo {author} {\bibfnamefont {A.~S.}\ \bibnamefont {S{\o}rensen}},\
  }\bibfield  {title} {\bibinfo {title} {Scalable {{Dissipative Preparation}}
  of {{Many-Body Entanglement}}},\ }\href
  {https://doi.org/10.1103/PhysRevLett.117.040501} {\bibfield  {journal}
  {\bibinfo  {journal} {Physical Review Letters}\ }\textbf {\bibinfo {volume}
  {117}},\ \bibinfo {pages} {040501} (\bibinfo {year} {2016})}\BibitemShut
  {NoStop}%
\bibitem [{\citenamefont {Xu}\ \emph {et~al.}(2019)\citenamefont {Xu},
  \citenamefont {Rao}, \citenamefont {Gui}, \citenamefont {Jin},\ and\
  \citenamefont {Hu}}]{xuCavitymediateddissipativeCoupling2019}%
  \BibitemOpen
  \bibfield  {author} {\bibinfo {author} {\bibfnamefont {P.-C.}\ \bibnamefont
  {Xu}}, \bibinfo {author} {\bibfnamefont {J.~W.}\ \bibnamefont {Rao}},
  \bibinfo {author} {\bibfnamefont {Y.~S.}\ \bibnamefont {Gui}}, \bibinfo
  {author} {\bibfnamefont {X.}~\bibnamefont {Jin}},\ and\ \bibinfo {author}
  {\bibfnamefont {C.-M.}\ \bibnamefont {Hu}},\ }\bibfield  {title} {\bibinfo
  {title} {Cavity-mediated dissipative coupling of distant magnetic moments:
  {{Theory}} and experiment},\ }\href
  {https://doi.org/10.1103/PhysRevB.100.094415} {\bibfield  {journal} {\bibinfo
   {journal} {Physical Review B}\ }\textbf {\bibinfo {volume} {100}},\ \bibinfo
  {pages} {094415} (\bibinfo {year} {2019})}\BibitemShut {NoStop}%
\bibitem [{\citenamefont {Barzanjeh}\ \emph {et~al.}(2017)\citenamefont
  {Barzanjeh}, \citenamefont {Wulf}, \citenamefont {Peruzzo}, \citenamefont
  {Kalaee}, \citenamefont {Dieterle}, \citenamefont {Painter},\ and\
  \citenamefont {Fink}}]{barzanjehMechanicalOnchipMicrowave2017}%
  \BibitemOpen
  \bibfield  {author} {\bibinfo {author} {\bibfnamefont {S.}~\bibnamefont
  {Barzanjeh}}, \bibinfo {author} {\bibfnamefont {M.}~\bibnamefont {Wulf}},
  \bibinfo {author} {\bibfnamefont {M.}~\bibnamefont {Peruzzo}}, \bibinfo
  {author} {\bibfnamefont {M.}~\bibnamefont {Kalaee}}, \bibinfo {author}
  {\bibfnamefont {P.~B.}\ \bibnamefont {Dieterle}}, \bibinfo {author}
  {\bibfnamefont {O.}~\bibnamefont {Painter}},\ and\ \bibinfo {author}
  {\bibfnamefont {J.~M.}\ \bibnamefont {Fink}},\ }\bibfield  {title} {\bibinfo
  {title} {Mechanical on-chip microwave circulator},\ }\href
  {https://doi.org/10.1038/s41467-017-01304-x} {\bibfield  {journal} {\bibinfo
  {journal} {Nature Communications}\ }\textbf {\bibinfo {volume} {8}},\
  \bibinfo {pages} {953} (\bibinfo {year} {2017})}\BibitemShut {NoStop}%
\bibitem [{\citenamefont {Bernier}\ \emph {et~al.}(2017)\citenamefont
  {Bernier}, \citenamefont {T{\'o}th}, \citenamefont {Koottandavida},
  \citenamefont {Ioannou}, \citenamefont {Malz}, \citenamefont {Nunnenkamp},
  \citenamefont {Feofanov},\ and\ \citenamefont
  {Kippenberg}}]{bernierNonreciprocalReconfigurableMicrowave2017}%
  \BibitemOpen
  \bibfield  {author} {\bibinfo {author} {\bibfnamefont {N.~R.}\ \bibnamefont
  {Bernier}}, \bibinfo {author} {\bibfnamefont {L.~D.}\ \bibnamefont
  {T{\'o}th}}, \bibinfo {author} {\bibfnamefont {A.}~\bibnamefont
  {Koottandavida}}, \bibinfo {author} {\bibfnamefont {M.~A.}\ \bibnamefont
  {Ioannou}}, \bibinfo {author} {\bibfnamefont {D.}~\bibnamefont {Malz}},
  \bibinfo {author} {\bibfnamefont {A.}~\bibnamefont {Nunnenkamp}}, \bibinfo
  {author} {\bibfnamefont {A.~K.}\ \bibnamefont {Feofanov}},\ and\ \bibinfo
  {author} {\bibfnamefont {T.~J.}\ \bibnamefont {Kippenberg}},\ }\bibfield
  {title} {\bibinfo {title} {Nonreciprocal reconfigurable microwave
  optomechanical circuit},\ }\href {https://doi.org/10.1038/s41467-017-00447-1}
  {\bibfield  {journal} {\bibinfo  {journal} {Nature Communications}\ }\textbf
  {\bibinfo {volume} {8}},\ \bibinfo {pages} {604} (\bibinfo {year}
  {2017})}\BibitemShut {NoStop}%
\bibitem [{\citenamefont {Caloz}\ \emph {et~al.}(2018)\citenamefont {Caloz},
  \citenamefont {Al{\`u}}, \citenamefont {Tretyakov}, \citenamefont {Sounas},
  \citenamefont {Achouri},\ and\ \citenamefont
  {{Deck-L{\'e}ger}}}]{calozElectromagneticNonreciprocity2018}%
  \BibitemOpen
  \bibfield  {author} {\bibinfo {author} {\bibfnamefont {C.}~\bibnamefont
  {Caloz}}, \bibinfo {author} {\bibfnamefont {A.}~\bibnamefont {Al{\`u}}},
  \bibinfo {author} {\bibfnamefont {S.}~\bibnamefont {Tretyakov}}, \bibinfo
  {author} {\bibfnamefont {D.}~\bibnamefont {Sounas}}, \bibinfo {author}
  {\bibfnamefont {K.}~\bibnamefont {Achouri}},\ and\ \bibinfo {author}
  {\bibfnamefont {Z.-L.}\ \bibnamefont {{Deck-L{\'e}ger}}},\ }\bibfield
  {title} {\bibinfo {title} {Electromagnetic {{Nonreciprocity}}},\ }\href
  {https://doi.org/10.1103/PhysRevApplied.10.047001} {\bibfield  {journal}
  {\bibinfo  {journal} {Physical Review Applied}\ }\textbf {\bibinfo {volume}
  {10}},\ \bibinfo {pages} {047001} (\bibinfo {year} {2018})}\BibitemShut
  {NoStop}%
\bibitem [{\citenamefont {Lecocq}\ \emph {et~al.}(2017)\citenamefont {Lecocq},
  \citenamefont {Ranzani}, \citenamefont {Peterson}, \citenamefont {Cicak},
  \citenamefont {Simmonds}, \citenamefont {Teufel},\ and\ \citenamefont
  {Aumentado}}]{lecocqNonreciprocalMicrowaveSignal2017}%
  \BibitemOpen
  \bibfield  {author} {\bibinfo {author} {\bibfnamefont {F.}~\bibnamefont
  {Lecocq}}, \bibinfo {author} {\bibfnamefont {L.}~\bibnamefont {Ranzani}},
  \bibinfo {author} {\bibfnamefont {G.~A.}\ \bibnamefont {Peterson}}, \bibinfo
  {author} {\bibfnamefont {K.}~\bibnamefont {Cicak}}, \bibinfo {author}
  {\bibfnamefont {R.~W.}\ \bibnamefont {Simmonds}}, \bibinfo {author}
  {\bibfnamefont {J.~D.}\ \bibnamefont {Teufel}},\ and\ \bibinfo {author}
  {\bibfnamefont {J.}~\bibnamefont {Aumentado}},\ }\bibfield  {title} {\bibinfo
  {title} {Nonreciprocal {{Microwave Signal Processing}} with a
  {{Field-Programmable Josephson Amplifier}}},\ }\href
  {https://doi.org/10.1103/PhysRevApplied.7.024028} {\bibfield  {journal}
  {\bibinfo  {journal} {Physical Review Applied}\ }\textbf {\bibinfo {volume}
  {7}},\ \bibinfo {pages} {024028} (\bibinfo {year} {2017})}\BibitemShut
  {NoStop}%
\bibitem [{\citenamefont {Metelmann}\ and\ \citenamefont
  {Clerk}(2015)}]{metelmannNonreciprocalPhotonTransmission2015}%
  \BibitemOpen
  \bibfield  {author} {\bibinfo {author} {\bibfnamefont {A.}~\bibnamefont
  {Metelmann}}\ and\ \bibinfo {author} {\bibfnamefont {A.~A.}\ \bibnamefont
  {Clerk}},\ }\bibfield  {title} {\bibinfo {title} {Nonreciprocal {{Photon
  Transmission}} and {{Amplification}} via {{Reservoir Engineering}}},\ }\href
  {https://doi.org/10.1103/PhysRevX.5.021025} {\bibfield  {journal} {\bibinfo
  {journal} {Physical Review X}\ }\textbf {\bibinfo {volume} {5}},\ \bibinfo
  {pages} {021025} (\bibinfo {year} {2015})}\BibitemShut {NoStop}%
\bibitem [{\citenamefont {Peterson}\ \emph {et~al.}(2017)\citenamefont
  {Peterson}, \citenamefont {Lecocq}, \citenamefont {Cicak}, \citenamefont
  {Simmonds}, \citenamefont {Aumentado},\ and\ \citenamefont
  {Teufel}}]{petersonDemonstrationEfficientNonreciprocity2017}%
  \BibitemOpen
  \bibfield  {author} {\bibinfo {author} {\bibfnamefont {G.~A.}\ \bibnamefont
  {Peterson}}, \bibinfo {author} {\bibfnamefont {F.}~\bibnamefont {Lecocq}},
  \bibinfo {author} {\bibfnamefont {K.}~\bibnamefont {Cicak}}, \bibinfo
  {author} {\bibfnamefont {R.~W.}\ \bibnamefont {Simmonds}}, \bibinfo {author}
  {\bibfnamefont {J.}~\bibnamefont {Aumentado}},\ and\ \bibinfo {author}
  {\bibfnamefont {J.~D.}\ \bibnamefont {Teufel}},\ }\bibfield  {title}
  {\bibinfo {title} {Demonstration of {{Efficient Nonreciprocity}} in a
  {{Microwave Optomechanical Circuit}}},\ }\href
  {https://doi.org/10.1103/PhysRevX.7.031001} {\bibfield  {journal} {\bibinfo
  {journal} {Physical Review X}\ }\textbf {\bibinfo {volume} {7}},\ \bibinfo
  {pages} {031001} (\bibinfo {year} {2017})}\BibitemShut {NoStop}%
\bibitem [{\citenamefont
  {Le~Boit{\'e}}(2020)}]{leboiteTheoreticalMethodsUltrastrong2020}%
  \BibitemOpen
  \bibfield  {author} {\bibinfo {author} {\bibfnamefont {A.}~\bibnamefont
  {Le~Boit{\'e}}},\ }\bibfield  {title} {\bibinfo {title} {Theoretical
  {{Methods}} for {{Ultrastrong Light}}--{{Matter Interactions}}},\ }\href
  {https://doi.org/10.1002/qute.201900140} {\bibfield  {journal} {\bibinfo
  {journal} {Advanced Quantum Technologies}\ }\textbf {\bibinfo {volume} {3}},\
  \bibinfo {pages} {1900140} (\bibinfo {year} {2020})}\BibitemShut {NoStop}%
\bibitem [{\citenamefont {Gardiner}\ and\ \citenamefont
  {Collett}(1985)}]{gardinerInputOutputDamped1985}%
  \BibitemOpen
  \bibfield  {author} {\bibinfo {author} {\bibfnamefont {C.~W.}\ \bibnamefont
  {Gardiner}}\ and\ \bibinfo {author} {\bibfnamefont {M.~J.}\ \bibnamefont
  {Collett}},\ }\bibfield  {title} {\bibinfo {title} {Input and output in
  damped quantum systems: {{Quantum}} stochastic differential equations and the
  master equation},\ }\href {https://doi.org/10.1103/PhysRevA.31.3761}
  {\bibfield  {journal} {\bibinfo  {journal} {Physical Review A}\ }\textbf
  {\bibinfo {volume} {31}},\ \bibinfo {pages} {3761} (\bibinfo {year}
  {1985})}\BibitemShut {NoStop}%
\bibitem [{\citenamefont {Yuan}\ \emph {et~al.}(2020)\citenamefont {Yuan},
  \citenamefont {Yu},\ and\ \citenamefont
  {Xiao}}]{yuanLoopTheoryInputoutput2020}%
  \BibitemOpen
  \bibfield  {author} {\bibinfo {author} {\bibfnamefont {H.~Y.}\ \bibnamefont
  {Yuan}}, \bibinfo {author} {\bibfnamefont {W.}~\bibnamefont {Yu}},\ and\
  \bibinfo {author} {\bibfnamefont {J.}~\bibnamefont {Xiao}},\ }\bibfield
  {title} {\bibinfo {title} {Loop theory for input-output problems in
  cavities},\ }\href {https://doi.org/10.1103/PhysRevA.101.043824} {\bibfield
  {journal} {\bibinfo  {journal} {Physical Review A}\ }\textbf {\bibinfo
  {volume} {101}},\ \bibinfo {pages} {043824} (\bibinfo {year}
  {2020})}\BibitemShut {NoStop}%
\bibitem [{\citenamefont {Bourhill}\ \emph {et~al.}(2023)\citenamefont
  {Bourhill}, \citenamefont {Yu}, \citenamefont {Vlaminck}, \citenamefont
  {Bauer}, \citenamefont {Ruoso},\ and\ \citenamefont
  {Castel}}]{bourhillGenerationCirculatingCavity2023}%
  \BibitemOpen
  \bibfield  {author} {\bibinfo {author} {\bibfnamefont {J.}~\bibnamefont
  {Bourhill}}, \bibinfo {author} {\bibfnamefont {W.}~\bibnamefont {Yu}},
  \bibinfo {author} {\bibfnamefont {V.}~\bibnamefont {Vlaminck}}, \bibinfo
  {author} {\bibfnamefont {G.~E.~W.}\ \bibnamefont {Bauer}}, \bibinfo {author}
  {\bibfnamefont {G.}~\bibnamefont {Ruoso}},\ and\ \bibinfo {author}
  {\bibfnamefont {V.}~\bibnamefont {Castel}},\ }\bibfield  {title} {\bibinfo
  {title} {Generation of {{Circulating Cavity Magnon Polaritons}}},\ }\href
  {https://doi.org/10.1103/PhysRevApplied.19.014030} {\bibfield  {journal}
  {\bibinfo  {journal} {Physical Review Applied}\ }\textbf {\bibinfo {volume}
  {19}},\ \bibinfo {pages} {014030} (\bibinfo {year} {2023})}\BibitemShut
  {NoStop}%
\bibitem [{\citenamefont {Zhang}\ \emph {et~al.}(2020)\citenamefont {Zhang},
  \citenamefont {Galda}, \citenamefont {Han}, \citenamefont {Jin},\ and\
  \citenamefont {Vinokur}}]{zhangBroadbandNonreciprocityEnabled2020}%
  \BibitemOpen
  \bibfield  {author} {\bibinfo {author} {\bibfnamefont {X.}~\bibnamefont
  {Zhang}}, \bibinfo {author} {\bibfnamefont {A.}~\bibnamefont {Galda}},
  \bibinfo {author} {\bibfnamefont {X.}~\bibnamefont {Han}}, \bibinfo {author}
  {\bibfnamefont {D.}~\bibnamefont {Jin}},\ and\ \bibinfo {author}
  {\bibfnamefont {V.~M.}\ \bibnamefont {Vinokur}},\ }\bibfield  {title}
  {\bibinfo {title} {Broadband {{Nonreciprocity Enabled}} by {{Strong
  Coupling}} of {{Magnons}} and {{Microwave Photons}}},\ }\href
  {https://doi.org/10.1103/PhysRevApplied.13.044039} {\bibfield  {journal}
  {\bibinfo  {journal} {Physical Review Applied}\ }\textbf {\bibinfo {volume}
  {13}},\ \bibinfo {pages} {044039} (\bibinfo {year} {2020})}\BibitemShut
  {NoStop}%
\bibitem [{\citenamefont {Ciuti}\ and\ \citenamefont
  {Carusotto}(2006)}]{ciutiInputoutputTheoryCavities2006}%
  \BibitemOpen
  \bibfield  {author} {\bibinfo {author} {\bibfnamefont {C.}~\bibnamefont
  {Ciuti}}\ and\ \bibinfo {author} {\bibfnamefont {I.}~\bibnamefont
  {Carusotto}},\ }\bibfield  {title} {\bibinfo {title} {Input-output theory of
  cavities in the ultrastrong coupling regime: {{The}} case of time-independent
  cavity parameters},\ }\href {https://doi.org/10.1103/PhysRevA.74.033811}
  {\bibfield  {journal} {\bibinfo  {journal} {Physical Review A}\ }\textbf
  {\bibinfo {volume} {74}},\ \bibinfo {pages} {033811} (\bibinfo {year}
  {2006})}\BibitemShut {NoStop}%
\bibitem [{\citenamefont {{Forn-D{\'i}az}}\ \emph {et~al.}(2019)\citenamefont
  {{Forn-D{\'i}az}}, \citenamefont {Lamata}, \citenamefont {Rico},
  \citenamefont {Kono},\ and\ \citenamefont
  {Solano}}]{forn-diazUltrastrongCouplingRegimes2019}%
  \BibitemOpen
  \bibfield  {author} {\bibinfo {author} {\bibfnamefont {P.}~\bibnamefont
  {{Forn-D{\'i}az}}}, \bibinfo {author} {\bibfnamefont {L.}~\bibnamefont
  {Lamata}}, \bibinfo {author} {\bibfnamefont {E.}~\bibnamefont {Rico}},
  \bibinfo {author} {\bibfnamefont {J.}~\bibnamefont {Kono}},\ and\ \bibinfo
  {author} {\bibfnamefont {E.}~\bibnamefont {Solano}},\ }\bibfield  {title}
  {\bibinfo {title} {Ultrastrong coupling regimes of light-matter
  interaction},\ }\href {https://doi.org/10.1103/RevModPhys.91.025005}
  {\bibfield  {journal} {\bibinfo  {journal} {Reviews of Modern Physics}\
  }\textbf {\bibinfo {volume} {91}},\ \bibinfo {pages} {025005} (\bibinfo
  {year} {2019})}\BibitemShut {NoStop}%
\bibitem [{\citenamefont {Frisk~Kockum}\ \emph {et~al.}(2019)\citenamefont
  {Frisk~Kockum}, \citenamefont {Miranowicz}, \citenamefont {De~Liberato},
  \citenamefont {Savasta},\ and\ \citenamefont
  {Nori}}]{friskkockumUltrastrongCouplingLight2019}%
  \BibitemOpen
  \bibfield  {author} {\bibinfo {author} {\bibfnamefont {A.}~\bibnamefont
  {Frisk~Kockum}}, \bibinfo {author} {\bibfnamefont {A.}~\bibnamefont
  {Miranowicz}}, \bibinfo {author} {\bibfnamefont {S.}~\bibnamefont
  {De~Liberato}}, \bibinfo {author} {\bibfnamefont {S.}~\bibnamefont
  {Savasta}},\ and\ \bibinfo {author} {\bibfnamefont {F.}~\bibnamefont
  {Nori}},\ }\bibfield  {title} {\bibinfo {title} {Ultrastrong coupling between
  light and matter},\ }\href {https://doi.org/10.1038/s42254-018-0006-2}
  {\bibfield  {journal} {\bibinfo  {journal} {Nature Reviews Physics}\ }\textbf
  {\bibinfo {volume} {1}},\ \bibinfo {pages} {19} (\bibinfo {year}
  {2019})}\BibitemShut {NoStop}%
\bibitem [{Note3()}]{Note3}%
  \BibitemOpen
  \bibinfo {note} {\begin {minipage}{\linewidth } \begin {flalign*} &\begin
  {aligned} &[\protect \hat {a}_i(t), \protect \hat {a}_j^\dagger (t)] = \delta
  _{ij}, \\ &[\protect \hat {b}_{\omega , i}(t), \protect \hat {b}_{\omega ',
  j}^\dagger (t)] = \delta _{ij}\delta (\omega - \omega '), \end {aligned}&
  \end {flalign*} \end {minipage} where $\delta _{ij}$ and $\delta (\omega -
  \omega ')$ are the Kronecker symbols.}\BibitemShut {Stop}%
\bibitem [{\citenamefont {Xiong}\ \emph {et~al.}(2010)\citenamefont {Xiong},
  \citenamefont {Zhang}, \citenamefont {Wang},\ and\ \citenamefont
  {Wu}}]{xiongExactNonMarkovianCavity2010}%
  \BibitemOpen
  \bibfield  {author} {\bibinfo {author} {\bibfnamefont {H.-N.}\ \bibnamefont
  {Xiong}}, \bibinfo {author} {\bibfnamefont {W.-M.}\ \bibnamefont {Zhang}},
  \bibinfo {author} {\bibfnamefont {X.}~\bibnamefont {Wang}},\ and\ \bibinfo
  {author} {\bibfnamefont {M.-H.}\ \bibnamefont {Wu}},\ }\bibfield  {title}
  {\bibinfo {title} {Exact non-{{Markovian}} cavity dynamics strongly coupled
  to a reservoir},\ }\href {https://doi.org/10.1103/PhysRevA.82.012105}
  {\bibfield  {journal} {\bibinfo  {journal} {Physical Review A}\ }\textbf
  {\bibinfo {volume} {82}},\ \bibinfo {pages} {012105} (\bibinfo {year}
  {2010})}\BibitemShut {NoStop}%
\bibitem [{\citenamefont {Moy}\ \emph {et~al.}(1999)\citenamefont {Moy},
  \citenamefont {Hope},\ and\ \citenamefont
  {Savage}}]{moyBornMarkovApproximations1999}%
  \BibitemOpen
  \bibfield  {author} {\bibinfo {author} {\bibfnamefont {G.~M.}\ \bibnamefont
  {Moy}}, \bibinfo {author} {\bibfnamefont {J.~J.}\ \bibnamefont {Hope}},\ and\
  \bibinfo {author} {\bibfnamefont {C.~M.}\ \bibnamefont {Savage}},\ }\bibfield
   {title} {\bibinfo {title} {The {{Born}} and {{Markov}} approximations for
  atom lasers},\ }\href {https://doi.org/10.1103/PhysRevA.59.667} {\bibfield
  {journal} {\bibinfo  {journal} {Physical Review A}\ }\textbf {\bibinfo
  {volume} {59}},\ \bibinfo {pages} {667} (\bibinfo {year} {1999})},\ \Eprint
  {https://arxiv.org/abs/quant-ph/9801046} {arxiv:quant-ph/9801046}
  \BibitemShut {NoStop}%
\bibitem [{\citenamefont {Bourhill}\ \emph {et~al.}(2020)\citenamefont
  {Bourhill}, \citenamefont {Castel}, \citenamefont {Manchec},\ and\
  \citenamefont {Cochet}}]{bourhillUniversalCharacterizationCavity2020}%
  \BibitemOpen
  \bibfield  {author} {\bibinfo {author} {\bibfnamefont {J.}~\bibnamefont
  {Bourhill}}, \bibinfo {author} {\bibfnamefont {V.}~\bibnamefont {Castel}},
  \bibinfo {author} {\bibfnamefont {A.}~\bibnamefont {Manchec}},\ and\ \bibinfo
  {author} {\bibfnamefont {G.}~\bibnamefont {Cochet}},\ }\bibfield  {title}
  {\bibinfo {title} {Universal characterization of cavity--magnon polariton
  coupling strength verified in modifiable microwave cavity},\ }\href
  {https://doi.org/10.1063/5.0006753} {\bibfield  {journal} {\bibinfo
  {journal} {Journal of Applied Physics}\ }\textbf {\bibinfo {volume} {128}},\
  \bibinfo {pages} {073904} (\bibinfo {year} {2020})}\BibitemShut {NoStop}%
\bibitem [{\citenamefont {Cao}\ and\ \citenamefont
  {Yan}(2019)}]{caoExceptionalMagneticSensitivity2019}%
  \BibitemOpen
  \bibfield  {author} {\bibinfo {author} {\bibfnamefont {Y.}~\bibnamefont
  {Cao}}\ and\ \bibinfo {author} {\bibfnamefont {P.}~\bibnamefont {Yan}},\
  }\bibfield  {title} {\bibinfo {title} {Exceptional magnetic sensitivity of
  {{P T}} -symmetric cavity magnon polaritons},\ }\href
  {https://doi.org/10.1103/PhysRevB.99.214415} {\bibfield  {journal} {\bibinfo
  {journal} {Physical Review B}\ }\textbf {\bibinfo {volume} {99}},\ \bibinfo
  {pages} {214415} (\bibinfo {year} {2019})}\BibitemShut {NoStop}%
\bibitem [{\citenamefont {Zhang}\ \emph {et~al.}(2019)\citenamefont {Zhang},
  \citenamefont {Ding}, \citenamefont {Zhou}, \citenamefont {Xu},\ and\
  \citenamefont {Jin}}]{zhangExperimentalObservationExceptional2019}%
  \BibitemOpen
  \bibfield  {author} {\bibinfo {author} {\bibfnamefont {X.}~\bibnamefont
  {Zhang}}, \bibinfo {author} {\bibfnamefont {K.}~\bibnamefont {Ding}},
  \bibinfo {author} {\bibfnamefont {X.}~\bibnamefont {Zhou}}, \bibinfo {author}
  {\bibfnamefont {J.}~\bibnamefont {Xu}},\ and\ \bibinfo {author}
  {\bibfnamefont {D.}~\bibnamefont {Jin}},\ }\bibfield  {title} {\bibinfo
  {title} {Experimental {{Observation}} of an {{Exceptional Surface}} in
  {{Synthetic Dimensions}} with {{Magnon Polaritons}}},\ }\href
  {https://doi.org/10.1103/PhysRevLett.123.237202} {\bibfield  {journal}
  {\bibinfo  {journal} {Physical Review Letters}\ }\textbf {\bibinfo {volume}
  {123}},\ \bibinfo {pages} {237202} (\bibinfo {year} {2019})}\BibitemShut
  {NoStop}%
\bibitem [{\citenamefont {Hurst}\ and\ \citenamefont
  {Flebus}(2022)}]{hurstNonHermitianPhysicsMagnetic2022}%
  \BibitemOpen
  \bibfield  {author} {\bibinfo {author} {\bibfnamefont {H.~M.}\ \bibnamefont
  {Hurst}}\ and\ \bibinfo {author} {\bibfnamefont {B.}~\bibnamefont {Flebus}},\
  }\bibfield  {title} {\bibinfo {title} {Non-{{Hermitian}} physics in magnetic
  systems},\ }\href {https://doi.org/10.1063/5.0124841} {\bibfield  {journal}
  {\bibinfo  {journal} {Journal of Applied Physics}\ }\textbf {\bibinfo
  {volume} {132}},\ \bibinfo {pages} {220902} (\bibinfo {year}
  {2022})}\BibitemShut {NoStop}%
\bibitem [{\citenamefont {Wang}\ \emph {et~al.}(2023)\citenamefont {Wang},
  \citenamefont {Huang}, \citenamefont {Qiu},\ and\ \citenamefont
  {Xiong}}]{wangNonreciprocalDoublecarrierFrequency2023}%
  \BibitemOpen
  \bibfield  {author} {\bibinfo {author} {\bibfnamefont {X.}~\bibnamefont
  {Wang}}, \bibinfo {author} {\bibfnamefont {K.-W.}\ \bibnamefont {Huang}},
  \bibinfo {author} {\bibfnamefont {Q.-Y.}\ \bibnamefont {Qiu}},\ and\ \bibinfo
  {author} {\bibfnamefont {H.}~\bibnamefont {Xiong}},\ }\bibfield  {title}
  {\bibinfo {title} {Nonreciprocal double-carrier frequency combs in cavity
  magnonics},\ }\href {https://doi.org/10.1016/j.chaos.2023.114137} {\bibfield
  {journal} {\bibinfo  {journal} {Chaos, Solitons \& Fractals}\ }\textbf
  {\bibinfo {volume} {176}},\ \bibinfo {pages} {114137} (\bibinfo {year}
  {2023})}\BibitemShut {NoStop}%
\bibitem [{\citenamefont {Agarwal}(1986)}]{agarwalGenerationPaircoherent1986}%
  \BibitemOpen
  \bibfield  {author} {\bibinfo {author} {\bibfnamefont {G.~S.}\ \bibnamefont
  {Agarwal}},\ }\bibfield  {title} {\bibinfo {title} {Generation of {{Pair
  Coherent States}} and {{Squeezing}} via the {{Competition}} of {{Four-Wave
  Mixing}} and {{Amplified Spontaneous Emission}}},\ }\href
  {https://doi.org/10.1103/PhysRevLett.57.827} {\bibfield  {journal} {\bibinfo
  {journal} {Physical Review Letters}\ }\textbf {\bibinfo {volume} {57}},\
  \bibinfo {pages} {827} (\bibinfo {year} {1986})}\BibitemShut {NoStop}%
\bibitem [{\citenamefont {Benatti}\ \emph {et~al.}(2003)\citenamefont
  {Benatti}, \citenamefont {Floreanini},\ and\ \citenamefont
  {Piani}}]{benattiEnvironmentInducedEntanglement2003}%
  \BibitemOpen
  \bibfield  {author} {\bibinfo {author} {\bibfnamefont {F.}~\bibnamefont
  {Benatti}}, \bibinfo {author} {\bibfnamefont {R.}~\bibnamefont
  {Floreanini}},\ and\ \bibinfo {author} {\bibfnamefont {M.}~\bibnamefont
  {Piani}},\ }\bibfield  {title} {\bibinfo {title} {Environment {{Induced
  Entanglement}} in {{Markovian Dissipative Dynamics}}},\ }\href
  {https://doi.org/10.1103/PhysRevLett.91.070402} {\bibfield  {journal}
  {\bibinfo  {journal} {Physical Review Letters}\ }\textbf {\bibinfo {volume}
  {91}},\ \bibinfo {pages} {070402} (\bibinfo {year} {2003})}\BibitemShut
  {NoStop}%
\bibitem [{\citenamefont {Kastoryano}\ \emph {et~al.}(2011)\citenamefont
  {Kastoryano}, \citenamefont {Reiter},\ and\ \citenamefont
  {S{\o}rensen}}]{kastoryanodissipativePreparationEntanglement2011}%
  \BibitemOpen
  \bibfield  {author} {\bibinfo {author} {\bibfnamefont {M.~J.}\ \bibnamefont
  {Kastoryano}}, \bibinfo {author} {\bibfnamefont {F.}~\bibnamefont {Reiter}},\
  and\ \bibinfo {author} {\bibfnamefont {A.~S.}\ \bibnamefont {S{\o}rensen}},\
  }\bibfield  {title} {\bibinfo {title} {Dissipative {{Preparation}} of
  {{Entanglement}} in {{Optical Cavities}}},\ }\href
  {https://doi.org/10.1103/PhysRevLett.106.090502} {\bibfield  {journal}
  {\bibinfo  {journal} {Physical Review Letters}\ }\textbf {\bibinfo {volume}
  {106}},\ \bibinfo {pages} {090502} (\bibinfo {year} {2011})}\BibitemShut
  {NoStop}%
\bibitem [{Note1()}]{Note1}%
  \BibitemOpen
  \bibinfo {note} {In the Heisenberg picture, an observable $\protect \hat {d}$
  satisfy: $\protect \dot {\protect \hat {d}} = - \protect \dfrac {i}{\hbar }
  [\protect \hat {d}, \protect \hat {\protect \mathcal {H}}]$}\BibitemShut
  {NoStop}%
\bibitem [{Note2()}]{Note2}%
  \BibitemOpen
  \bibinfo {note} {Note that $\DOTSI \intop \ilimits@ \limits _{+\infty }^t
  \protect \mathrm {d}t' \protect \, f(t') = - \DOTSI \intop \ilimits@ \limits
  _t^{+\infty } \protect \mathrm {d}t' \protect \, f(t')$, where $f(t')$ is a
  time-dependent function.}\BibitemShut {Stop}%
\end{thebibliography}%

\end{document}